\documentclass{aims}

\usepackage{amsmath}
\usepackage{amssymb}
\usepackage{amsthm}
\usepackage{bm}
\usepackage{color}   
\usepackage{subfigure}
\newcommand{\etal}{\textit{et al}. }

 \usepackage{setspace} 
  \usepackage{paralist}
  \usepackage{graphics} 
  \usepackage{epsfig} 
\usepackage{graphicx}  
\usepackage{epstopdf}
\usepackage{textcomp}

 \usepackage[colorlinks=true]{hyperref}
\usepackage{url}

\hypersetup{urlcolor=blue, citecolor=red}

\def\currentvolume{X}
 \def\currentissue{X}
  \def\currentyear{200X}
   \def\currentmonth{XX}
    \def\ppages{X--XX}
     \def\DOI{10.3934/xx.xx.xx.xx}

\usepackage[british,UKenglish,USenglish,english,american]{babel}
\usepackage{ragged2e,array,booktabs}
\usepackage[noadjust]{cite}
\usepackage{enumerate}
\usepackage[left = 3.00cm, right = 3.00cm, top = 3.00cm, bottom=3.00cm]{geometry}
\usepackage{longtable}
\usepackage[T1]{fontenc}

\usepackage{times}

\usepackage[table]{xcolor}

\title[The future of cancer treatments]{Prospect for application of mathematical models in combination cancer treatments}

	\author[Joseph Malinzi, Kevin Bosire Basita, Sara Padidar and Henry A. Adeola]{}


 \email{josephmalinzi1@gmail.com}
 \email{kevinb@aims.ac.za }
 \email{henry.adeola@uct.ac.za}
\email{sarapadidar@gmail.com}

\thanks{$^*$ Corresponding author: Henry A. Adeola , Email address: henry.adeola@uct.ac.za}

\begin{document}
\maketitle
\centerline{\scshape Joseph Malinzi}
\medskip
{\footnotesize
 \centerline{Department of Mathematics, Faculty of Science and Engineering, University of Eswatini, Private Bage 4, Kwaluseni, Eswatini }
 \centerline{Institute of Systems Science, Durban University of Technology, Durban 4000, South Africa}	 
} 

\medskip

\centerline{\scshape Kevin Bosire Basita}
\medskip
{\footnotesize
 \centerline{Department of Mathematics, Faculty of Applied Science and Technology, Technical University of Kenya, Haile Selassie Avenue, Nairobi, Kenya}

}
\medskip 

\centerline{\scshape Sara Padidar}
\medskip
{\footnotesize
 \centerline{Department of Biological Sciences, Faculty of Science and Engineering, University of Eswatini, Private Bage 4, Kwaluseni, Eswatini}

}

\medskip

\centerline{\scshape Henry A. Adeola$^*$}
\medskip
{\footnotesize
 \centerline{Department of Medicine, Division of Dermatology, Faculty of Health Sciences and Groote Schuur Hospital,}

 \centerline{ University of Cape Town, Cape Town 7925, South Africa }
}

\bigskip
\bigskip

{\centerline{Corresponding author: Henry A. Adeola; Email address: henry.adeola@uct.ac.za }}

\bigskip


\begin{abstract}
	
The long-term efficacy of targeted therapeutics for cancer treatment can be significantly limited by the type of therapy and development of drug resistance, inter alia.  Experimental studies indicate that the factors enhancing acquisition of drug resistance in cancer cells include cell heterogeneity, drug target alteration, drug inactivation, DNA damage repair, drug efflux, cell death inhibition, as well as microenvironmental adaptations to targeted therapy, among others. Combination cancer therapies (CCTs) are employed to overcome these molecular and pathophysiological bottlenecks and improve the overall survival of cancer patients. CCTs often utilize multiple combinatorial modes of action and thus potentially constitute a promising approach to overcome drug resistance. Considering the colossal cost, human effort, time and ethical issues involved in clinical drug trials and basic medical research,  mathematical modeling and analysis can potentially contribute immensely to the discovery of better cancer treatment regimens. In this article, we review mathematical models on CCTs developed thus far for cancer management. Open questions are highlighted and plausible combinations are discussed based on the level of toxicity, drug resistance, survival benefits, preclinical trials and other side effects.

\end{abstract}

\section{Introduction}\label{s1}
Cancer is the second leading cause of death globally, with 70\% of cancer mortality from low and middle income countries \cite{WHO2018}. To date, cancer treatment has heavily relied on surgery, often followed by chemotherapy and/or radiotherapy \cite{hu2016recent,beil2002sequencing,benzekry2017contributions}. Recent advances with the use of virological and immunological agents including nanotechnology delivery methods have increased our arsenal in treating this complex disease \cite{heath2008nanotechnology}. Given the high costs involved in treating cancer, and the continued development of better and less toxic treatment approaches, mathematical modeling provides a valuable and low-cost tool for researchers by predicting treatment outcomes and identifying potential best combinations to therapeutic approaches with minimal adverse effects.

The use of mathematical modeling in understanding a variety of disease states, including cancer, has been well documented \cite{byrne2010dissecting,araujo2004history}. Computational biology and mathematical modeling have enabled researchers to develop new treatments, from screening potential candidate compounds \cite{barbolosi2016computational} and building and identifying chemical structures that will best fit target sites \cite{byrne2010dissecting,kuang2016introduction} to predicting the outcome of novel combinations of drug regimens \cite{chappell2015mathematical,malinzi2017modelling,malinzi2018enhancement,goldie1988mathematical,2000single,kim2015quantitative,friedman2018combination,dingli2006mathematical}. This paper reviews a novel use of mathematical modeling to enable researchers to predict the treatment outcomes of combined therapeutic strategies in the treatment of cancers.


Benign cancers and locally confined malignant cells can often be treated by a single treatment strategy such as surgery or radiation therapy alone. In some cases, a combination approach is employed such as the use of (neo) adjuvant radiotherapy or chemotherapy before or after surgery to reduce the size of a tumour, thereby increasing chances of complete surgical resection and/or destroying of any post-surgical residual cancer cells \cite{hsieh2018adjuvant,trimble1993neoadjuvant}. Employing combined therapeutic strategies is essential to many malignant tumours where a single approach (e.g. surgery) is not always suitable, for instance in bladder cancer or non-small cell lung cancer that may not be operable. 

Although mono-therapeutic strategies claim a large share in the medical world for the treatment of several cancer types, they are considered less effective in comparison with the combination therapy approach \cite{ottolino2010intelligent}. Most standard mono-therapeutic strategies (e.g. chemotherapy alone) destroy all the actively proliferating cells non-selectively, ultimately causing the annihilation of both healthy and cancerous cells \cite{ottolino2010intelligent}. In addition, chemotherapy is known to produce toxic effects on the patients with a myriad of side effects, threat to life, suppressing bone marrow cells and increasing vulnerability to infection \cite{ottolino2010intelligent,chan2016chemotherapy}. Hence, there is an overarching demand for tumour-targeted therapies, that are minimally toxic especially in the era of precision and personalized medicine.

Combination cancer treatments hold great appeals in boosting the efficacy of anticancer drugs, enhancing apoptosis, suppressing tumour growth and decreasing cancer stem cell population \cite{malinzi2019mathematical}. Although chemotherapy-based combination therapy could be toxic, the toxicity can be greatly reduced when different pathways are targeted \cite{mokhtari2017combination,malinzi2019mathematical}. Moreover, dosing for these agents works in an additive manner, thereby lowering the cumulative amount of therapeutic dosage needed in each individual drug  \cite{mokhtari2017combination}.

Several advances have been made in tumour profiling, and deep sequencing has revealed driver mutations as well as novel targets for the development of new cancer drugs \cite{dry2016looking,doroshow2017design}. There have been remarkable achievements on combination therapy in recent years with much attention focussing on most effective combinations that are likely to produce significant effects on the tumour. However, this journey has not been smooth with the process of designing plausible combinations facing numerous challenges. Current combination cancer therapy presumes that better results are obtained by amalgamating therapeutic drugs at their maximal tolerated doses \cite{lopez2017combine}. Nevertheless, differences in the pharmacokinetics of single drugs bring about inconsistency in the delivery of synergistic drug ratios to tumour cells. Compounded toxicity observed in combination treatment protocols calls for application of suboptimal dosages to counter the effects. Additionally, there are challenges regarding the identification of the best combination strategies and the complications involved in the combination therapy \cite{lopez2017combine}.

Upon the selection of a reasonably designed combination, the process of its early clinical development is complicated; thus requiring attention to detail \cite{lopez2017combine}. Implementation of combination strategies is influenced by several issues including toxicity, pharmacokinetic and pharmacodynamic interactions, proper timing of resistance development and the identification of robust biomarker in predicting the response. Clearly, the questions of how to implement the combination, when to implement the combination and to whom it is implemented for are the major problems that face the clinical community during decelopment of treatment regimens \cite{lopez2017combine}.

Production of new anticancer drugs takes a long time, is risky, and requires costly medical experimentation; and thus mathematical and computational approaches are needed to supplement clinical trials in making predictions. For this reason, efficient mathematical approaches are needed to try and understand different mechanisms of cancer activities and effective combination therapy for cancer at a relatively affordable cost. Many Mathematical models have been formulated to determine effective combination cancer therapies. Nonetheless, the cure for cancer is still far from reach, with complexities associated with therapeutic combinations remaining a major impediment.  

\begin{figure}
\centering 
\includegraphics[scale=0.5]{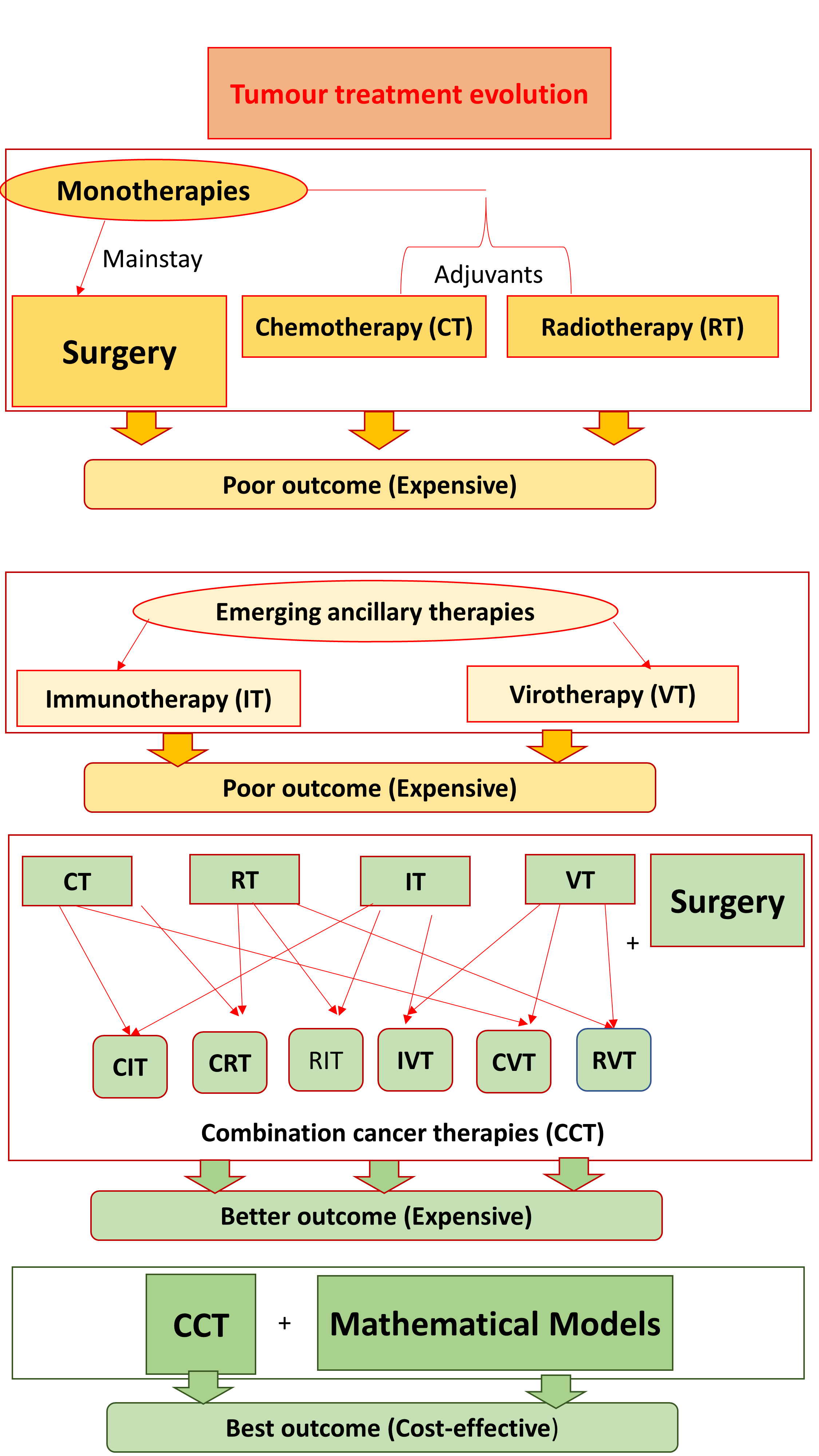}
\caption{A schematic diagram showing the evolution of cancer treatments. It depicts better therapeutic results with the use combination cancer strategies and that mathematical models play an integral role.}
\label{fig:flow_diagram}
\end{figure}

Here, we carry out a critical review of published mathematical models that have been used to predict the outcome of combination therapy strategies. The therapeutic strategies reviewed are chemoimmunotherapy, chemovirothrapy, chemoradiotherapy, radioimmunotherapy, immunovirotherapy, radiovirotherapy and targeted therapies (see Figure \ref{fig:flow_diagram}: which shows the evolution of cancer treatments. The figure further depicts that better therapeutic outcomes are achieved when combination cancer strategies are used. It as well depicts that mathematical models shall play an integral role in deciding on strategies that are viable). Existing open problems are discussed and we propose plausible combinations of treatment strategies to treat a variety of cancers.

\section{Mathematical models for combination cancer therapy}\label{section:mathematical models}
\subsection{Models of Chemoimmunotherapy}
Cancer chemotherapeutic drugs normally disrupt pathways necessary for the growth and survival of tumours. However, they are often limited due to the development of drug resistance and the damages exerted to normal cells of the host tissue \cite{emens2010chemoimmunotherapy}. Recent years have focused on host-tumour interactions  and it has been established that host-tumour interactions play a pivotal role in determining clinical course and the outcomes of the treatments of human malignancies \cite{emens2010chemoimmunotherapy}. Several mathematical models have been constructed in an attempt to understand  tumour-immune interactions in depth and determine the elements in the immune system that play a critical role in responding to immunotherapy \cite{de2009mathematical}.  Several studies, \cite{de2007mathematical,adam1997general,chaplain2006mathematical,de1985macrophage, arciero2004mathematical,de2006mixed,de2005validated}, have attempted to understand the effects of immune modulation by using mathematical approaches. 
 
Modern cancer treatment methods rely on the ability of certain cancers to trigger the immune response. Incorporation of the immune component in the formulation of mathematical models has played a key role in the clinically observed phenomena, for instance, tumour dormancy, unchecked growth of tumours and oscillations in tumour size \cite{isaeva2009different}. The first attempt to illustrate the immunotherapy related effects in a suitable ordinary differential equations (ODEs) model was given by Kirschner and Panetta \cite{kirschner1998modeling}. The study considered interleukin (IL-2) along with adaptive cellular immunotherapy (ACI) by developing dynamical equations explaining external inflow of  both IL-2 and immune cells. 

Chappell \etal \cite{chappell2015mathematical} presented a high-level abstraction mathematical model that explored the interaction of immune cells and tumour cells, where they further explored the merits of combining immunotherapeutic agents with chemotherapy and radiotherapy. The model was tested and numerical simulations were carried out using data similar to that shown by Deng \etal \cite{deng2014irradiation}. Numerical results showed that tumour mass reduced significantly when radiotherapy and immunotherapy are used in combination whereas single therapy recorded no significant decrease in tumour mass. Further, a combination of radiotherapy and immunotherapy leads to an increase in the number of T-cells that are activated compared to single therapies.

Another mathemtical model, comprising of ODEs, focussed on the interplay between NK and CD8$^+$ T cells with different varieties of tumour cell lines \cite{de2003mathematical}. The results found that ligand transduced cells trigger enough immune response to put tumour growth under control, whereas control-transduced tumour cells evade immunity.

In a more recent study, de Pillis \etal \cite{de2009mathematical} extended their earlier study \cite{de2003mathematical} by incoporating the latest research on baseline NK and activated CD8$^+$ T-cells in both healthy donors and cancer patients. The model was qualitatively analyzed and supplemented with numerical simulations of chemoimmunotherapy and vaccine therapy. The results of the study intimated that, depending on the efficacy of the CD8$^+$ T-cells, chemoimmunotherapy is a likely to succesfully control tumour cells in body tissue. 
 
Rodrigues et al. \cite{rodrigues2019mathematical} developed a simple ODE model to study the dynamics of chemoimmunotherapy of chronic lymphocytic leukaemia. Their model analysis showed that the application of chemotherapy paves way for immunotherapy in a way that it decreases the number of cancer cells thus reducing the tumour density on which immune cells acts upon. It was established that there is a minimum number of immune cells needed to be transplanted in adoptive treatment so as to ensure a complete cure or remission. Nonetheless, more studies are needed to model adoptive cellular immunotherapy.

Pang \etal \cite{pang2016mathematical} discussed single chemotherapy, immunotherapy and mixed treatments as well as conditions triggering tumour eradication. They analysed the effects of least effective concentration and the half-life of the therapeutic drugs where they found that better results can be achieved if the half-life of the drug is extended. Further, the impact of drug resistance on therapeutic results was considered and a mathematical model explaining the cause of chemotherapeutic failure that uses a single drug was proposed. In the end, numerical simulations showed that chemoimmunotherapy is likely to achieve a better treatment effect. However, despite the reported achievement, tumour cells are continously becoming resistant to numerous structurally and mechanistically unrelated drugs, limiting the effectiveness of chemoimmunotherapy. Thus, determining how to effectively amalgamate those treatment modes and design optimal dosage combination regimens deserves further research.

\subsection{Models of chemovirotherapy}
The use of viruses as targeted therapy in cancer originated in the early 20th century with numerous viruses already tested in humans and experimental animals \cite{vaha2007oncolytic}. Oncolytic viruses (OVs) are a heterogeneous group of low or non pathogenic viruses employed in the treatment of cancers, taking advantage of their ability to selectively infect and lyse cancer cells without damaging normal cells \cite{guo2008oncolytic,fukuhara2016oncolytic}. Many viruses, both genetically engineered and naturally occurring, are currently being researched as oncolytic cancer virotherapeutic agents \cite{guo2008oncolytic}. Examples of such viruses include vaccinia virus, measles virus, reovirus, adenovirus and vesicular stomatitis virus \cite{li2012oncolytic,yu2009oncolytic}. During carcinogenesis, tumour cells are able to gain genetic and cellular changes that interferes with the host immunosurveillance capabilities \cite{mokhtari2017combination}. Such targeted, tumour-specific properties make oncolytic virotherapy, either solo or in combination (e.g. chemovirotherapy), an attractive therapy for cancer.

In recent years, chemovirotherapy has been shown to enhance therapeutic effects that are far from reach when each therapy is used independently \cite{malinzi2019mathematical,ottolino2010intelligent,nguyen2014chemotherapy,ungerechts2010mantle,malinzi2017modelling,ulasov2009combination}. Ottolio-Perry \etal \cite{ottolino2010intelligent} reviewed the combination of oncolytic viruses with the existing treatment modalities with a view of helping investigators make informed decisions regarding the clinical development of these products. In another study, Nguyen \etal \cite{nguyen2014chemotherapy} highlighted the mechanisms that guarantee the successful combination of  oncolytic viruses with other therapeutic drugs where, however, he concluded that the success of these combinations is subject to several factors including the type of oncolytic virus used, type of cancer targeted, dosage and the timing.

There is a wealth of mathematical models that have been developed in an attempt to comprehend as well as characterize viral dynamics \cite{friedman2006glioma,friedman2003analysis, wein2003validation, wu2001modeling,tao2005competitive,eftimie2018tumour,JPH14}. Not many mathematical models have, nonetheless-thus far, been developed and studied to investigate the combination of chemotherapeutic drugs and viruses. Malinzi \etal \cite{malinzi2017modelling} constructed and analysed a model of parabolic non-linear partial differential equations (PDEs) to examine the spatiotemporal dynamics of tumour cells under the treatment of chemovirotherapy. The proposed model consisted compartments of uninfected and infected tumour cells densities, a free virus, and a chemotherapeutic drug. It was shown that the success of virotherapy is highly determined by the virus infection rate and burst size. Further, travelling wave solutions revealed that tumour diffusivity and growth rate are critical parameters during chemovirotherapy treatment and numerical simulations confirmed that combining chemotherapeutic drugs with oncolytic viruses is more effective compared to either monotherapies.

In an effort to investigate the enhancement of chemotherapy by oncolytic viruses, Malinzi \etal \cite{malinzi2018enhancement} proposed an ODE  mathematical model and an optimal control problem for chemovirotherapy. The model described the interplays between tumour cells, immune response and a treatment combination by combining virotherapy and chemotherapy. Stability analysis showed that tumours could grow to their maximum capacity given that there is no form of treatment. It was shown that chemotherapy alone is likely to clear tumour cells on the condition that the efficacy of the drug is more than the intrinsic growth rate of the tumour. The combined effects of oncolytic virotherapy and chemotherapy were evaluated using sensitivity analysis of the model parameters. Optimal control simulations revealed that the half of the maximum tolerated dozes (MTDs) for chemotherapy and virotherapy optimise treatment outcomes. Sensitivity analysis and numerical simulations affirmed that the success of chemovirotherapy depends on virus burst size, the rate of viral infection and the dose. Moreover, the right dose of chemotherapy required to produce effective results in unison with virotherapy has been a concern for both clinicians and mathematicians. 

Malinzi \cite{malinzi2019mathematical} proposed a mathematical model for chemovirotherapy where he considered three-drug infusion methods and compared their efficacies, carried out mathematical analysis to forecast the outcomes of the OVs in combination with chemotherapy and also compared the efficacy of each single treatment modality. The model was formulated based on two aspects, that is, model without delay and with delay. Numerical simulations for both models were carried out, where it was shown that if the tumour burst is big, then regardless of the drug infusion method, chemovirotherapy is more effective than any of the single treatments. Simulations further indicated that the success of chemovirotherapy depends on virus burst size, the rate of viral infection and the dose. Further, simulations showed that incorporating delays in the system increased the time within which tumour clearance occurred in body tissue.
Despite this study, there still exists open questions which need further investigation, for example, the model does not address the quantity of dose required to completely eradicate the tumour and at what specific point in the tumour should the drug be infused to have a greater effect. Interestingly, ascertaining the optimal dosing schedule is still a challenge.

\subsection{Models of Chemoradiotherapy}
The paradigm of the interplay between radiation and chemotherapy was first conceptualized by Steel and Peckham in 1979 \cite{steel1979exploitable} and was later summarized by Seiwert \etal \cite{grassberger2016methodologies}. Two of the main adavantages for the use of chemoradiotherapy, assuming there is no interaction between the two, is to use cytotoxic drugs that will be used to tackle the disease, not within the radiation field and to irradiate seclusion sites \cite{steel1979exploitable}.

In a study by Goldie and Coldman \cite{goldie1988mathematical}, a stochastic model for alternating radiation and chemotherapy was proposed. Their model was based on the earliest approaches for combination therapy where they used three compartments, that is, stem, differentiation and end cells in modeling tumour growth. Their model incorporated chemo-resistant, radio-resistant and extra parameters to measure cells with joint resistance. They found that a regimen treatment with three doses of chemotherapy and radiotherapy acting in an alternating manner is powerful compared to sequential treatment under the same conditions because it suppresses subpopulation resistant responsible for treatment failure.

In another study by Beil and Wein \cite{beil2002sequencing}, a mathematical model was formulated with the aim of determining the best way to sequence the three standard therapies for cancer, that is, radiotherapy, chemotherapy and surgery. Differential equations were used to model the growth of tumours and its metastases with an underlying assumption that primary and metastatic tumours behave identically. The model aimed to show which combination achieved higher curative probability: surgery followed by chemotherapy and radiotherapy (SCR ), Surgery followed by radiotherapy and chemotherapy (SRC) and surgery followed by radiotherapy then chemotherapy and radiotherapy (SRCR) given that the primary tumour is big enough or that the metastatic cancer population outnumbers the primary tumour. Their model included metastatic cancer originating from dormant cells and angiogenesis, which eventually led to the interactions between primary tumour compartment and metastatic compartment and thus the emergence of optimality sequences.

Ergun \etal \cite{ergun2003optimal} used a similar approach to find optimal scheduling and doses of angiogenic inhibitors and radiotherapy that maximize the elimination of a primary tumour. The model was  analysed using two compartments of tumour cells and vascular endothelial cells and the damage by radiation was modelled in accordance with the linear-quadratic which widely determines fractionation schedules in radiobiology society. It was shown that optimal amalgamation treatment regimen happens when antiangiogenetic therapy is constantly increased in order to maintain an optimal tumour endothelial cells ratio, and the fraction sizes of radiation keep changing with treatment so as to maximize the probability of tumour control. However, the model was solely concerned with the primary tumour and did not include the antimetastatic effect of antiangiogenic treatment,  which might inhibit metastatic growth during radiotherapy if it is applied methodically.  Further, the model did not take into consideration the fact that radiation may cause vascular endothelial growth factor (VEGF) expression, which attenuates the destruction of endothelial cells by radiotherapy.

In a study by Salari \etal \cite{salari2015mathematical} it was shown that the addition of chemotherapy drug and its mode of action makes optimal fractionation regimen change for concurrent chemoradiotherapy. They contend that the introduction of chemotherapeutic drugs containing cytotoxic effects only does not hamper optimal fractionation regimen when there is no concurrent chemotherapy whereas a radiosensitization agent has the capacity to tamper with the optimality choice of fraction size. They also show that the emergence of optimal non-uniform fractionation may result if a drug exhibits both radiosensitization and independent cytotoxic properties \cite{medina2012can}.

Ghaffari \etal \cite{ghaffari2016mixed} proposed an ODE model that considered chemotherapy and radiotherapy for metastatic cancer. They examined the interplay between immune and cancer cells, with chemotherapy regarded as a predator on both normal and cancer cells. The model aimed at understanding the specific system dynamics and as well guide the creation of combination therapies. It was shown that a decaying dosage protocol is weaker compared to a constant dosage protocol in eliminating small metastases at a given time. 

Marcu \etal \cite{marcu2006scheduling} simulated ciplatin-radiotherapy treatment with emphasis on time sequencing and combined treatment scheduling of the drug and radiation. They examined different time courses of cisplatin and  radiotherapy for advanced head and neck cancer. It was agreed that to discover the true synergistic effect between cisplatin and radiotherapy reported in the literature, they would have to incorporate a proper radiosensitizing term. They further indicated that by administering cisplatin on a weekly basis would result to enhanced therapeutic effects compared to daily doses and that giving cisplatin in cross proximity in time to the radiation fractions would definitely result to maximal combined effect.

Barazzuol \etal \cite{barazzuol2010mathematical} formulated a mathematical model of radiotherapy  in  the  treatment  of  glioblastoma  where they used a linear quadratic model and investigated the additional effect of temozolomide in two simplified  scenarios. Their main aim was to address the question of whether the benefit of temozolomide is drawn from its radiosensitizing properties in the concurrent phase or from the  additional  cell  kill  of  the  adjuvant  phase.  They  concluded  that  the  model  presuming  only  radiosensitization  fits  the  observed  survival  curves  more  closely  than  assuming  independent cytotoxicity. 

Powathil \etal \cite{powathil2012modelling} presented a multiscale mathematical model which relied  on cellular automata  to  examine  the  spatial  distribution  and microenvironment  of  a  cell  population during treatment. A log cell-kill model was used to estimate chemotherapeutic effects whereas  the  linear-quadratic  model was used in the case of radiation effects. They applied their model to show a clinical application  to two known and one hypothetical treatment regimen for esophageal cancer,  where they concluded that their proposed alternative regimen was more effective. Several mathematical models allow broad disruptions regarding the concepts of pure radio-sensitization and independent chemotherapy cell kill \cite{jones1999inclusion,jones2005potential,plataniotis2008use,plataniotis2014assessment,moraru2014radiation,durante2015modeling,hartley2010radiobiological,meade2013revised,meade2013revising}.

\subsection{Models of Radioimmunotherapy}
Successful implementation of radioimmunotherapy for cancer treatment  has proven  to be significantly more  difficult  than  was expected initially \cite{sgouros1992plasmapheresis,epenetos1986limitations,order1990radiolabeled,begent1990antibody,sharkey1990biological}. This is, in part, as a result of highly  diverse, complex and interrelated biological and  physical factors that are considered when designing a successful protocol \cite{sgouros1992plasmapheresis}. By utilizing  mathematical models incorporating  plasma antibody pharmacokinetics, extravasation and interstitial transport and the  antibody-antigen interaction, Fujimori \etal  and van Osdol   \etal \cite{fujimori1989modeling,fujimori1990modeling,fujimori1991integrated,van1991analysis} investigated the connection between a  variety  of tumour and  antibody specific parameters and the microscopic distribution of antibody and absorbed dose within a tumour \cite{sgouros1992plasmapheresis}.

O'Donoghue \etal \cite{2000single} proposed a Uniform  tumour Dosimetry mathematical model that compared single-dose and fractionated radioimmunotherapy. The model compared large single administrations (LSAs) with rapid fractionation (RF) administered in small quantities within a short period of the time interval. It was indicated that for homogeneous absorbed dose   distributions throughout tumours, LSAs are predicted to  give a greater possibility of tumour cure than rapidly fractionated treatments of the same marrow toxicity. If dose  distributions are heterogeneous, RF may have a therapeutic advantage, depending on how tumour  uptake varies from one fraction to another.

Kumar \cite{kumar2010mathematical} proposed a mathematical model for radioimmunotherapy where they formulated radiotherapy dose distributions with regard to optimizing tumour cure probability (TCP) where the rate of dose distribution was assumed to be high enough to enable the delivery of dose distribution instantaneously. Numerical results showed that the tumour cells density and dose distribution are not sensitive to other various functional forms of tumour parameters. The results of this study revealed that the immune response plays a critical role during cancer treatment.

Flux \etal \cite{flux1997three} described a dosimetry 3-D mathematical model that quantified the absorption of dose distribution as a result of administration of an intralesional radiolabeled monoclonal antibody which allowed the distribution of spatial and temporal heterogeneity of radionuclide without a calibration scan. The model was tested by using a set of registered patient data where dose profiles and histograms of the dose volume were produced. It was established that 3-D dose distribution was significantly non-uniform. Initially, intralesional infusion results suggested that the model offered a means of determining how the absorbed dose was distributed within a tumour.

Serre \etal \cite{serre2016mathematical} suggested a discrete-time pharmacodynamic mathematical model of the amalgamation of radiotherapy and immune checkpoint inhibitors: PD1-PDL1 axis and the CTLA4. The model shows how a growing tumour triggers and inhibits tumour-immune responses and describes the effects of irradiation. The model's ability to predict pharmacodynamic endpoints was justified in retrospective by examining that it could explain clearly data obtained from experimental studies, which took into consideration the  combination of radiotherapy and immune checkpoint inhibitors.
They concluded that different designs in silico could be compared by simulating Kaplan-Meier curves and mathematical tools could be used to partially automate optimized protocols.

More work is needed to focus on integrating the developed mathematical methods into clinical practice that will heavily deal with managing drug and radiation-induced toxicities during the process of optimization. Further, combining radioimmunotherapy with other treatment regimens including targeted agents and  metronomic chemotherapy is a subject to be explored in future research. Morestill, dose effect relationships should be examined and a pharmacokinetic model integrated to clearly shade light on the relationship between the experimental results and the actual dosing protocols.

\subsection{Models of Immunovirotherapy}
Viruses have been engineered genetically to promote the expression of pro-inflammatory cytokines and co-stimulatory molecules which, upon release, instigates a targeted immune response on the tumour \cite{melcher2011thunder,cheema2014immunovirotherapy,hardcastle2017immunovirotherapy,cerullo2012oncolytic,wares2015treatment}. Several treatment approaches combining viral oncolysis with production of immunostimulatory molecules and dendritic cell (DC) injections are currently under construction with the aim of bettering treatment outcomes \cite{bagheri2011dynamical,huang2010therapeutic, wares2015treatment}. Several studies have been carried out to investigate the outcome of immunovirotherapy on myeloma \cite{dingli2009dynamics}, to establish how initial conditions hamper the efficacy of OV therapy on melanoma \cite{rommelfanger2012dynamics} and determine factors inhibiting and enhancing the transmission of oncolytic viruses via a tumour site \cite{mok2009mathematical}. Other recent studies focussing on the interaction of immune cells with DC vaccines include \cite{depillis2013model,pappalardo2014induction, dritschel2018mathematical}.

Lai and Friedman \cite{friedman2018combination}  considered a combination therapy where one drug was used as a vaccine to activate dendritic cells so as to instigate more T cells for the purpose of invading the tumour while the other drug is a checkpoint inhibitor which suppresses cancer cells. They developed a mathematical model in the form of PDEs to address the question of whether combining treatment of two drugs administered at certain levels is better than using a treatment of one drug with almost twice the dosage level. The model describes the interaction of dendritic cells, cancer cells, cytokines IL-I2 and IL-2, CD8$^+$ cells $(T_8)$, GM-CFC-secreting cancer cells (GVAX) and anti-PD-1. The concept of synergy between drugs was introduced and a synergy map was developed suggesting the proportion in which drugs should be administered so as to realize the maximum tumour volume to be reduced under the restriction of a maximum tolerated dose. Results showed that combining GVAX and anti-PD-1 in suitable quantities could consequentially reduce the growth of a tumour. More studies need to be carried out to explore the effects of treatment with GM-CFC-secreting vaccine (GVAX) and anti-PD-1 drug.

To date, oncolytic virotherapy still posses an exercise in population dynamics where the interplays between the viruses, tumour cells and the immune system components play a major role in determining treatment outcomes
\cite{wu2004analysis,berg2019vitro,wodarz2003gene,wodarz2012complex,bajzer2008modeling,biesecker2010optimization,dingli2006mathematical,dingli2009dynamics,rommelfanger2012dynamics,friedman2006glioma}. Several mathematical models have been formulated to describe the outcome of such interactions \cite{wu2004analysis,berg2019vitro,wodarz2003gene,wodarz2012complex,bajzer2008modeling,biesecker2010optimization,dingli2006mathematical,rommelfanger2012dynamics}.
A study that addresses the combined effects of viral oncolysis and T-cell-mediated oncolysis was carried out in \cite{kim2015quantitative} wherein a mathematical model of virotherapy that induces release of cytokine IL-12 and co-stimulatory molecule 4-1BB ligand was developed. It was found that  whereas viral oncolysis is important in reducing the tumour burden, increased stimulation of cytotoxic T cells leads to a short term reduction in tumour size, but a faster relapse. In addition, it was found that amalgamations of specialist viruses expressing either IL-12 or 4-1BBL might initially act more potently against tumours than a general virus that simultaneously expresses both, but the advantage is likely not large enough to replace treatment using the generalist virus.

A mathematical model incorporating cytokine and co-stimulatory molecule expressing OVs in combination with DC injections was studied by Wares \etal \cite{wares2015treatment}. The aim of this study was to investigate the effect of varrying doses of OV and DC injections during tumour treatment. It was shown using simulations that treatment of a tumour with immunostimulatory OVs first followed by a sequence of DC injections is more effective than alternating OV and DC injections. It was concluded that the efficacy of a dosing strategy depends on the ordering of oncolytic adenoviruses and dendritic cells, temporal tumour spacing and dosages chosen. This model was later extended by Gevertz and Wares \cite{gevertz2018developing} in which different techniques were used; such as information criteria analyses, sensitivity analyses and a parameter sloppiness analysis and it was discovered that it is possible to reduce their model to one variable and three parameters and still be fitted to the data. They argued that reduction of the model to minimal form allows for tractability of the system in the face of parametric uncertainty.

Timalsina \etal \cite{timalsina2017mathematical} proposed a PDE model with the aim of studying tumour virotherapy and mediated immunity. The model incorporated adaptive and innate immunity responses with interactions in tumour cells, OVs and immune system in a territory containing a boundary in motion. They assumed that all tumour cells exhibit homogeneity and viruses follow a diffusion process. They used the results obtained to examine tumour development and provide insight into crucial aspects of virotherapy such as dependence of the efficacy on vital parameters and the delay in the adaptive immunity.

Jenner \etal \cite{jenner2018modelling} suggested a mathematical model with a view of reproducing the results for tumour growth synonymous to results produced experimentally during treatment with an adenovirus virus. The model was fitted to data using parameters from the literature in order to reduce the degrees of freedom of the model. Optimization of the model was done using  a least-squares non-linear fitting algorithm `lsqnonlin' in MATLAB. The model indicates that decreasing  APC stimulation and raising helper T cell stimulation is likely to improve treatment. Nonetheless, it is still imperative to establish the range of these parameters when the absolute elimination of tumour cells occur.

Mahasa \etal \cite{mahasa2017oncolytic} presented a mathematical model that elucidated the interplays between oncolytic viruses, tumour cells, normal cells and the antitumoural and antiviral immune responses. The model consisted of delay defferential equations (DDEs) with one discrete delay indicating the time required to trigger a tumour-specific immune response. The model aimed to predict effects of antitumoural and antiviral immune responses in the presence of oncolytic viruses, the response of the tumour to the oncolytic viral infections and the oncolytic virus tumour-specificity responsible for maximizing the reduction of the tumour while concurrently ensuring that the toxicity on the normal tissue encircling tumour cells are minimized.

Cassidy and Humphries \cite{cassidy2018mathematical} formulated and analysed a mathematical model of tumour growth under oncolytic virotherapy and immunotherapy. The model explicitly included heterogeneity in tumour propagation speed by considering the cell cycle duration as a random variable. The interrelationship between the estimated number of cells surviving in the cell cycle and tumour eradication was established by the help of linear stability analysis. The results of their analysis showed that an increase in immune recruitment acts synergistically to lyse tumour cells during oncolytic virotherapy. Nonetheless, they noted that their model was limited by the fact that it oversimplified tumour-immune interplays. 

\subsection{Models of Radiovirotherapy}

The interaction dynamics of oncolytic viruses with tumour cells and immune responses are highly intricate \cite{dingli2006mathematical,wu2001modeling,wein2003validation,wodarz2001viruses,wodarz2003gene}. Determining the outcome of radiovirotherapy is dependent on understanding the complex
interactions between the various components involved during treatment with radiovirotherapy \cite{dingli2006mathematical}. Thus, modeling the kinetics of virotherapy could act as an impetus for the development of improved therapeutic protocols as well as gaining further insight on the outcome of such therapies \cite{dingli2006mathematical,wu2001modeling,wein2003validation,wodarz2001viruses,wodarz2003gene}. 

A study by Dingli \etal \cite{dingli2006mathematical} focused on attenuated strains of measles virus that highly infect tumour cells because of their high expression of CD46, a property that gives viruses an advantage to attach and enter into the target cells. They proposed and analysed a mathematical model that took into account population dynamics that encapsulated crucial components of radiovirotherapy. Analysis of the model included determining and establishing existence of corresponding equilibria related to complete cure, partial cure, and treatment failure. Through numerical simulations, they analyzed the influence of factors, in the form of parameters, that determine the outcome of radiovirotherapy treatment. Morestill, they evaluated relevant therapeutic scenarios for effective radiovirotherapy. The study showed that radiotherapy is more effective when used synergistically with oncolytic virotherapy.

Tao and Guo \cite{tao2007free} developed a PDE model describing cancer radiovirotherapy, which is a generalization of the existing ODE models given in \cite{dingli2006mathematical,friedman2003analysis}. The model was developed as a moving boundary problem. Global existence and uniqueness of solutions to the model was proved, and a new explicit parameter condition corresponding to treatment success was determined. By modeling combined action of virotherapy and radiotherapy, they aimed to design the possible optimal therapy strategies. Numerical experiments verified that radiovirotherapy is more effective compared to monotherapy. To explore possible optimal therapy strategies, a number of numerical simulations were further carried out. The results suggested that there is an optimal timing of radio-iodine administration and an optimal dose of the radioactive iodide, which needs to be tested by experimental data.

\subsection{Models of targeted therapies}
Targeted therapies involve the use of novel cancer drugs that interact with specific molecular targets that are pertinent to cancer progression \cite{NCI,abbott2006mathematical}. Examples of targeted therapies include; protein-kinase inhibitors: Imatinib (Glivec\textregistered)-used for the treatment of leukemia, Dasatanib (Sprycel\textregistered)-used for treating  prostate cancer, and Temsirolimus (Torisel\textregistered)-used for treating several cancer types, monoclonal antibodies: Bevacizumab (Avastin \textregistered)-for treating colon cancer, and proteasome inhibitors: Bortezomib (Valcade\textregistered)-for treating myeloma and leukemia \cite{abbott2006mathematical}. Dozens of other targeted therapies for treating different types of cancer do exist and have already been approved for use whereas others are undergoing clinical trials \cite{bozic2013evolutionary}. Single targeted therapy drugs with short-lived treatment effects often fail [116]. The alternative strategy previously proposed is to synergistically use a combination of targeted therapy drugs which employ different pathways, or to use targeted therapies together with other treatment methods. For example, with immunotherapy; a great number of preclinical and clinical trials have revealed that there is a synergistic antitumor effect with the use of targeted therapy with immunotherapy \cite{yu2019combination}.

The development or improvement of these targeted therapies requires a molecular understanding of cancer pathogenesis and a characterisation of the interplay between tumour cells and therapeutic agents. Nonetheless, not many attempts have been made to mathematically investigate the dynamic relationships between cancerous cells and targeted therapeutic agents; moreover only a few mathematical studies exist, thus far, for the prediction of the outcome of combination treatment with targeted therapies. A comprehensive review of mathematical models for targeted therapy can be found in Abbott and Michor \cite{abbott2006mathematical}. We complement these with a review of some mathematical studies on targeted therapies and combination treatments involving targeted therapies.

Green \etal \cite{green2001mathematical} proposed a compartmental mathematical model for antibody-targeted therapy of colorectal cancer, using clinical trial data from bio-distribution of antibodies against carcinoembryonic antigen. The study indicated the most significant parameters that determine antibody localization are the affinity for the antibody, the flow of the antibody through the tumour and the rate of elimination of the antibody from the tumour.

Shen \etal \cite{shen2020biphasic} developed a mathematical biphasic model that accurately described the cell-drug response based on an analysis of targeted inhibition of colorectal cancer cell lines. Their model used three kinetic parameters: ratio of target-specific inhibition, F1, potency of target-specific inhibition, Kd1, and potency of off-target toxicity, Kd2 to describe the drug response where the determination of these kinetic parameters are valuable in predicting effective combination targeted therapy for multi-driver cancer cells. The model and the mechanistic insights give a new mechanistic perspective and mathematical tool for predicting effective combination targeted blockades against multi-driver cancer cells.

Stochastic differential equations (SDEs) have been employed to model the stochastic evolution of resistance of therapeutic drugs. Sun \etal \cite{sun2016mathematical} proposed a stochastic model by using a set of SDEs to examine the dynamics of drug sensitive cells, drug-resistant cells and new metastatic cells. They used clinical data to validate their model and predicted distinct patterns of drug dose-dependent synergy for two different sets of drug combinations in the treatment of melanoma. This approach was expected to enhance the study of effective and robust cancer
drugs.

Kozlowska \etal \cite{kozlowska2018mathematical} presented a comprehensive stochastic mathematical model and simulator approach that describes platinum resistance and standard of care therapy in high-grade serous ovarian cancer (HGSOC). Their study used post treatment clinical data, including 18F-FDG-PET/CT images, to accurately predict the model parameters and simulate `virtual patients with HGSOC'. The model revealed that the treatment responses exhibited by these virtual patients was in line with those of real-life patients with HGSOC. They utilized their approach to evaluate survival benefits of combination therapies containing up to six drugs targeting platinum resistance mechanisms where they indicated that combining standard of care with a drug specifically targeting the most dominant resistance sub-population resulted in a significant survival benefit.

Jarrett \etal \cite{jarrett2019experimentally} constructed a mathematical model that evaluated the combination of trastuzumab-paclitaxel therapies for drug interaction relations on the basis of order, timing and quantity of the drug dosage.  Their model was based on time-resolved microscopy data that captured changes in vitro cell confluence in response to the combination of paclitaxel and trastazumub treatment. The model showed increased synergy for treatment regimens where trastuzumab was administered before paclitaxel and indicated that trastuzumab accelerates the cytotoxicity of paclitaxel. 

Bozic \etal \cite{bozic2013evolutionary} proposed a mathematical model to predict the effect of combining targeted therapies. Their model was based on data obtained from 20 melanoma patients. Their model simulation results revealed that combination treatment with two targeted therapy drugs is far more effective than using a single drug; and those drugs should simultaneously be administered for optimal results.

Owen \etal \cite{owen2011mathematical} constructed a spatiotemporal mathematical model to determine the outcome of combining macrophage-based hypoxia-targeted gene therapy with chemotherapy. Model simulation results showed that combining conventional drugs and macrophage based targeted therapies would work synergistically and result to better anti-cancer effects than the individual effects from each of the therapies.

  Arciero \etal \cite{arciero2004mathematical} presented a mathematical model  where they considered a novel treatment strategy known as small interfering RNA (siRNA) therapy in which the treatment suppresses TGF-$\beta$ production by targeting the mRNA codes for TGF-$\beta$, and thus reducing the presence and effect of TGF-$\beta$ in tumour cells. The model predicts conditions within which siRNA treatment can be successful in transformation of TGF-$\beta$ producing tumours to either non-producing or producing a small value of TGF-$\beta$ tumours.
  
  Tang \etal \cite{tang2013target} introduced a novel model, called TIMMA (Target Inhibition inference using Maximization and Minimization Averaging) with the aim of demonstrating its feasibility in systematic investigation of the model predictions using kinome-wide single and pairwise siRNA knock-down experiments. The model used functional data on drugs for its construction and target combination predictions. The validation of TIMMA results using systematic siRNA-mediated silencing of the selected targets and their pairwise combinations showed an increased ability to identify both druggable kinase targets and synergistic interactions. The model was applied to case studies in MCF-7 and MDA-MB-231 breast cancer and BxPC-3 pancreatic cancer cells where it was confirmed that TIMMA-predicted kinase targets are essential for tumor survival, either individually or in combination. Further, it was shown that the construction of model algorithm resulted in significantly better predictive accuracy and computational efficiency in comparison with an existing algorithmic solution.
  
  In another study, Tang \etal \cite{tang2019network} applied network pharmacology model to predict synergistic drug combinations. Kinome-wide drug-target profiles and gene expression data were used to pinpoint a synergistic target interaction between Aurora B and ZAK kinase inhibition that resulted to an enhanced growth inhibition and cytotoxicity. They employed stochastic simulation algorithm (SSA) to model signalling pathways, implement and understand the mechanisms of action of the identified target interactions. An MDA-MB-231 signaling network was constructed using dynamic modeling approach to simulate the effect of perturbing the genes of interests on the cell viability. The observed similarity of the model predictions to the experimental observations indicates that the constructed signaling network consists of key protein-protein interactions that may mediate the synergistic and antagonistic relationship of Aurora B with ZAK and CSF1R, respectively.
 
 Araujo \etal  \cite{araujo2005mathematical} presented a mathematical model to investigate combination therapy in which multiple nodes in a signaling cascade pathway are simultaneously targeted with specific inhibitors. The model suggested that the attenuation of biochemical signals is significantly enhanced when several upstream processes are inhibited, and that this weakening is most pronounced in signals downstream of serially connected targets. 
 
 Komarova and Dominik \cite{komarova2005drug} developed a mathematical framework to examine the principles underlying the emergence and prevention of resistance in the treatment of cancers with targeted drugs. They considered a stochastic dynamical system which took into account turnover rate of tumor cells and the rate of generating resistant mutants.The model was applied to chronic myeloid leukemia where it was suggested that combining three targeted drugs with different specificities is likely to overcome the problem of resistance.
 
 Charusanti \etal \cite{charusanti2004mathematical} presented a mathematical model describing several signalling events in chronic myeloid leukemia (CML) cells. Their study examined dynamic effects of the drug STI-571 (Glivec) on the autophosphorylation of the BCR-ABL oncoprotein and subsequent signalling through the Crkl pathway, and they used simulations to predict a minimal concentration for drug effectiveness. The model was fitted to mouse Bcr-Abl autophosphorylation
data where it is revealed that cellular drug clearance mechanisms such as drug efflux significantly reduces the effectiveness of Glivec in blast crisis cells. Additionally it is stipulated that these resistance mechanisms might be present from the onset of disease.
  
Other mathematical models of combination cancer therapy include  \cite{shen2002model,callahan2013two, chakwizira2018mathematical, monjazeb2013combined, radunskaya2018mathematical,nazari2018mathematical,wang2016optimal,malinzi2015mathematical,spring2019illuminating,rihan2014delay,oke2018optimal,mamat2013mathematical,lai2018modeling,joshi2009immunotherapies,kosinsky2018radiation,sgouros1992plasmapheresis,rihan2019optimal,akman2018optimal,ratajczyk2018optimal,sharma2016analysis,ledzewicz2014optimal,nazari2015finite,parra2013mathematical,rodrigues2013mathematical,villasana2010modeling,chareyron2009mixed,ledzewicz2008optimal,d2009optimal,powathil2007mathematical,imbs2018revisiting,tao2008mathematical,hadjiandreou2014mathematical,su2016optimal,kim2012modeling,bajzer2008modeling}. Bajzer \etal \cite{bajzer2008modeling} modelled cancer virotherapy with recombinant measles viruses where the interactions of the tumour and virus were based on the already established biology.  Hadjiandreou and Mitsis \cite{hadjiandreou2014mathematical} developed a mathematical model that used Gompertz growth law and pharmacokinetic-pharmacodynamic approach to model the effects of drugs on on tumour progression and designed optimal therapeutic patterns. Su \etal \cite{su2016optimal} formulated a mathematical model of tumour therapy with oncolytic viruses and MEK inhibitor.

\clearpage 
\begin{longtable}{p{3.1cm}p{5cm}p{5.00cm}p{1.5cm}}
\caption{A summary of merits and demerits associated with the use of certain forms of combination therapies.} \\
\hline
\label{tab:comparisonsofthedifferenttherapies}

	Therapies & Merits& Demerits & References\\
	\hline
Chemoimmunotherapy&
	\begin{itemize}
	\item  Application of low-dose chemotherapy with immune-stimulating vaccination significantly lengthens survival both in a prophylactic and therapeutic setting.
	\item Improved progression-free survival and the overall survival of physically fit patients with previously untreated symptomatic chronic lymphocytic leukemia.
	\end{itemize} 
	&
	\begin{itemize}
	\item Although chemoimmunotherapy  in  animal models is well established, it is extremely hard to transfer this  knowledge  into clinical practice as there are several hurdles that  must be overcome.
	\item Side effects include secondary malignancies, myelo- and immunosuppression which are likely to be the onset of myeloid malignancies (acute myeloid leukemia and myelodysplastic syndrome), hematological non-hematological toxicities.
	\end{itemize}
	  
&\cite{chappell2015mathematical,deng2014irradiation,shrivastava2018cisplatin,maletzki2019chemo,medina2012can,brown2016chemoimmunotherapy,hallek2010addition,galluzzi2014classification,sasse2007chemoimmunotherapy} \\
\hline

Chemovirotherapy &  
\begin{itemize}
\item Several studies have shown that combining chemotherapy with oncolytic viruses results to enhanced therapeutic effects.
\item  Benefits include significant improvement of overall survival, enhanced efficacy and decreased tumour volume. 
\end{itemize}

& 
\begin{itemize}
\item Although this combination has proved promising, it is however limited by the side effects such as viremia and cardiac toxicity.
\end{itemize}
&\cite{malinzi2017modelling,malinzi2018enhancement,zhang2014there,ottolino2010intelligent,binz2015chemovirotherapy,malinzi2019mathematical,advani1998enhancement,chung2002use,blank2002replication}\\
		\hline

	Chemoradiotherapy & 
	\begin{itemize}
	\item  Cisplatin chemoradiotherapy (CRT) delays disease recurrence and boosts survival in contrast to radiotherapy (RT) alone in women with stage IIIB squamous cell carcinoma of the cervix.
	\item CCRT has a survival advantage compared with RT alone in patients between 65 and 72 years.
	\item Enhances local tumour control in the treatment of cervical cancer.
\item Concurrent chemoradiotherapy has been shown to reduce the risk of cancer recurring by 50\% in patients having advanced stage disease regional spread.
\end{itemize}	  &  \begin{itemize}
\item Moderate to acute gastrointestinal tract effects in the form of bleeding proctitis due to telangiectasia and ulceration, which has been poorly reported in the literature.
	\item  Increased incidences of tumour recurrence, gastrointestinal and severe  hematologic complications.
	\item  Has  low overall  survival rate and late toxicity  
\end{itemize} &\cite{beil2002sequencing,goldie1988mathematical,steel1979exploitable,eifel2006chemoradiotherapy,neuner2009chemoradiotherapy,shitara2009chemoradiotherapy,  baxi2016trends,ottolino2010intelligent}.\\
	 \hline

		Radioimmunotherapy &
		\begin{itemize}
		\item Radioimmunotherapy treatment regimen has been used in the treatment of haematological malignancies with a fractionated dosing regimen achieving remarkable regression of bulky masses with non-Hodgkin lymphoma.
		\item Studies on other aggressive non-Hodgkin lymphoma have indicated promising efficacy of anti-CD20 radioimmunotherapy and anti-CD22 radioimmunotherapy
			 reduced haematologic toxicity, improved survival benefits and improved efficacy. 
		\end{itemize} 
		&
	\begin{itemize}
	\item Limited by hematologic toxicity, secondary cancers and of myelodysplastic syndrome.
	\item There is considerable evidence indicating that the overall risk for Radioimmunotherapy is no higher than chemotherapy.
	\item  Other demerits include the development of drug resistance as well as being costly.
	\item Convincing of the oncohematologist community to incorporate radioimmunotherapy would require designing of randomized clinical trials and stratification of patients for response to radioimmunotherapy. Clinical efficacy of radioimmunotherapy applications in solid tumours remain limited.
	\end{itemize}

		&\cite{2000single,sharkey2011cancer,kraeber2016radioimmunotherapy,gopal2003high}\\
		
		\hline
		
Immunovirotherapy&
		\begin{itemize}
		\item Selective virus replication in rapidly proliferating cells.  
		\item Directs tumour microenvironment towards immune response.
		\item Enhances survival. 
		\item OVs produces pathogens that are responsible for instigating as well as enhancing anti-tumour immunity. Combining OVs with immunostimulatory cytokines initiates an additional anti-tumourimmune response with apoptosis of tumour cells occurring due to suppressive properties of immune cells. 
\end{itemize}		    & 
	\begin{itemize}
	\item Genetic mutations allow for the development of resistance of tumour cells to OVs. Moreover tumour cells can learn to avoid adaptive immunity via immunoediting. 
	\item Viruses can trigger anti-viral immune responses. 
\end{itemize}		
  
&\cite{gevertz2018developing,jenner2018modelling,melcher2011thunder,cheema2014immunovirotherapy,hardcastle2017immunovirotherapy,cerullo2012oncolytic,wares2015treatment}\\
\hline

		Radiovirotherapy&
		\begin{itemize}
		\item In the treatment of prostate cancer, it is reported that MV-NIS results to significant tumour regression as well as highly significant prolongation of survival in animals.
		\item In addition, administration of iodine-131 enhances the antitumour effect of MV-NIS virotherapy.
		\item Several HSV1 mutants act synergistically with radiation therapy 83-85, although in some experiments the impact of radiation is essentially additive
\end{itemize}		    &
	\begin{itemize}
	\item Side effects including gastrointestinal toxicity and weight loss.
	\item The efficacy of radiovirotherapy can be limited by several factors in thyroid tumour treatment.
	\item The biodistribution of the virus, its propagation in the tumour through numerous replication cycles and the strength of the sodium/iodide  symporter expression can be inadequate or heterogeneous and thus it may affect an effective therapeutic response.
\end{itemize}		
  
&\cite{dingli2006mathematical,msaouel2009noninvasive,li2011oncolytic,dingli2004image,ottolino2010intelligent,touchefeu2012radiovirotherapy,chiocca2002oncolytic,advani1998enhancement,chung2002use,blank2002replication,spear2000cytotoxicity,jorgensen2001ionizing}\\
\hline

	Targeted therapies &
		\begin{itemize}
		\item Foundation of precision medicine. 
		\item Drug combinations targeting  different pathways can be used. 
		\item Attenuation of biochemical signals is highly enhanced. 
		\item Imatinib (Glivec\textregistered), the first targeted therapy to be approved by the FDA, has proven to be successful in the treatment of chronic myeloid leukaemia.
\end{itemize}		    &
	\begin{itemize}
	\item Side effects include gastrointestinal toxicity and weight loss.
	\item Cancer cells can eventually become resistant to targeted therapies. 
	\item Drugs for some targets are hard to develop.  
\end{itemize}		
  
& \cite{NCI,abbott2006mathematical,yu2019combination,araujo2005mathematical}\\
\hline

		\end{longtable}

\begin{longtable}{p{3.1cm}p{5cm}p{5.00cm}p{1.5cm}}
\caption{Examples of existing software for implementing mathematical models reviewed in this paper.}
\label{tab:software}
 \\
\hline  
 Software & Packages & Functionality & References \\
 \hline
	Matlab &Contains several differential equation solvers, for example, ode23 and pdepe and tool boxes, for example, jLab; used for data analysis. & Popular system for data visualization and carrying out numerical simulations of differential equations.&\cite{matlab}\\
	 \hline
	Maple & Differential equation packages like dsolve and PDEtools.
	 
	& Powerful software for symbolic computation and qualitative analysis. &\cite{maple} \\
\hline
Mathematica & Contains functions such as Dsolve and NDsolve. 
	&Powerful symbolic computation; offers numerical evaluation, optimization and visualization of a very wide range of numerical functions. &\cite{Mathematica}\\
	\hline

		Python & Contains packages like Scilab; a numerical computational package that can be used for handling ODEs and PDEs.  &
	
		Open source software and a high level, numerically oriented programing language.  &\cite{Python}\\
		
 \hline
		SageMath & Rich in several packages such as desolve for solving differential equations. 
		 & Open source software covering numerous areas including differential calculus, algebra, numerical analysis, calculus and statistics. &\cite{Sage}

\\
\hline
	
Maxima &  desolve function for solving ordinary differential equations. & Used for symbolic computation and solving differential equations.  & \cite{Maxima}\\
		\hline
GNU Octave & ODE solver; LSODE. & Powerful open source software with built-in 2D/3D plotting and visualization tools. &\cite{Octave}\\
		\hline
Mathcad & Contains functions like odesolve and rkfixed.  & Contains equation-solving packages which can be used to plot functions, solve blocks of equations, perform curve-fitting operations as well as solving differential equations using iterative methods.&\cite{MathCad}\\
\hline
Julia & Contains solvers like OrdinaryDiffEq.jl and LSODA.jl.   & Open source software designed for high performance computing; can be used for determining numerical simulations stochastic ODEs and PDEs among others.   &\cite{Julia}\\
\hline 
R & Contains ODE solvers like deSolve.    & Open source software designed mainly for statistical computing and graphics; can be used for carrying out numerical simulations of ODEs and PDEs.   &\cite{R}\\
\hline

 COPASI& Contains solvers like Hybrid LSODA/SSA and LSODAR. & An open source software that can be used for solving mathematical models of biological processes. & \cite{COPASI}\\
\hline

CompuSyn & & Analysis of combination data. &\cite{CompuSyn}\\
\hline

 CalcuSyn && Analyzes combined drug effects and is able to automatically quantify phenomena such as synergism and inhibition. & \cite{CalcuSyn}\\
 \hline

 TIMMA-R & R packages. & Predicts the effects of drug combinations based on their binary drug-target interactions and single-drug sensitivity profiles in a given cancer sample.&\cite{he2015timma}\\
 \hline

 Synergyfinder & Web application. & Allows for pre-processing, analyzing and visualising pairwise drug combinations in an interactive manner. &\cite{ianevski2020synergyfinder}\\
 \hline

Combenefit & & Enables for the analysis, quantification and visualization of drug combination effects in terms of synergy and/or antagonism. &\cite{di2016combenefit}\\
 \hline

  URDME& & Simulates stochastic reaction-diffusion processes on arbitrary meshes. &\cite{URDME}\\
  \hline
 
  COMSOL& & Solves highly nonlinear differential equations. &\cite{COMSOL}\\
  \hline

\end{longtable}
  	\begin{longtable}{p{3.1cm}p{5cm}p{5.00cm}p{1.5cm}}
  	\caption{Existing database for combinations of cancer therapy.}
  	\label{tab:database}
 \\
\hline 
		
		\hline

		Database & Potential applications & Website & References\\
		
		\hline
	
DrugComb & A data portal for integrative cancer drug combinations. & \url{https://drugcomb.fimm.fi/}&\cite{zagidullin2019drugcomb}\\

\hline 

		NCI ALMANAC &Provides data showing how well pairs of FDA-approved cancer drugs kill tumor cells from the NCI-60 Human Tumor Cell Lines. & \url{https://dtp.cancer.gov/ncialmanac/initializePage.do}&\cite{combination}\\
		\hline
	
		DrugCombDB& Comprehensive database dedicated to collecting drug amalgamations
from various data sources. &\url{http://drugcombdb.denglab.org/main}&\cite{liu2020drugcombdb}\\
		\hline
	
		 CMTTdb& Focuses on the key information for randomized clinical trials (RCTs), such as patient selection criteria, tumor conditions, therapy modes (single or combined therapy).&\url{https://www.biosino.org/CMTTdb/}&\cite{bai2019cmttdb}\\
		 
		 \hline
		
		 CANCROX& Provides important information about cancer treatment, drug combinations, genes and types of cancer.&\url{http://cancrox.gmb.bio.br/view/index.php}&\cite{de2019cancrox}\\
		 
		 \hline
		
		 OncoRx& Detecting drug interactions with chemotherapy regimens.&\url{https://onco-informatics.com/}&\cite{yap2010cancer}\\
		 
		 \hline
		
		 SynToxProfiler&Promotes the prioritization of drug combinations based on integrated analysis of synergy, efficacy and toxicity profiles. &\url{https://syntoxprofiler.fimm.fi/stp/eerrnKuW9DeMSZ3rclWtuLyrBpHbUvN+WDUAXDxZLUg=/} &\cite{ianevski2020syntoxprofiler}\\
		 \hline
		
		 ReDO-DB & Focuses exclusively on the potential use of licensed non-cancer medications as sources of new cancer therapeutics. & \url{http://www.redo-project.org/db/} &\cite{pantziarka2018redo_db}\\
		 \hline
		
		 CancerDR & Identification of mutations in cancer genes (targets) as well as natural variations in targets. & \url{https://webs.iiitd.edu.in/raghava/cancerdr/#thumb} &\cite{kumar2013cancerdr}
	
		\end{longtable}

\section{Discussion and conclusion}
Combination cancer therapy strategies have fast become the established method for oncologists to treat cancer, and has the potential to counter treatment side effects and resistance to drugs. There is much concern over the toxicity and effectiveness of combining therapeutic strategies. Ideally, there are several factors that necessitate the selection of a certain cancer combination therapy strategy such as improvement in the quality of life of the patient, nonoverlapping safety profiles, reduction of drug resistance and preclinical evidence of synergy between the interacting therapeutic agent. However, developing novel combinations of therapeutic strategies is time-consuming and expensive and depends on a myriad of factors unique to each cancer. Moreover, there are still several questions that need to be answered pertaining to combination cancer therapy. These include determining the optimal dosage combinations and investigating the side effets of using the different combination treatments. Mathematics can help to answer such questions by allowing researchers to model therapeutic options (including new, repurposed and old drugs) and predict their treatment and toxicity outcomes before embarking on lengthy and expensive trials. Scientists use a range of methods, tools and software to analyse mathematical models; the choice depends on the problem being solved and the model type. Mathematical models in the form of differential equations are normally analysed using qualitative methods; to investigate the long-term characteristics of the model solutions \cite{preziosi2003cancer,kelley2010theory,brauer2012mathematical}. Model analysis is usually complemented with numerical and computational simulations using several methods implemented by a number of programing languages and software \cite{Polyanin,Software}. Some examples of tools and software that are used in analysing mathematical models reviewed in this paper and tools that can be used for users to upload and analyse input data are summarised in Table \ref{tab:software}. We as well provide some existing databases for combinations of cancer therapy in Table \ref{tab:database}. 

Complete drug development takes more than 10 years with huge costs ranging from USD 12 million to USD 12 billion. Moreover, only a small fraction of drugs that enter clinical trials are eventually approved \cite{dimasi2003price}. This makes it imperative to devise methods of reducing the cost of drug development and one such way is through mathematical modeling. Drug developers, researchers, clinicians and other bio-scientists can use model simulations and results from mathematical analyses to assimilate complex mechanisms of cancer therapy dynamics thus leading  the development of more effective treatments.

In Section \ref{section:mathematical models}, we reviewed mathematical models for a number of combination cancer therapy strategies developed thus far. In doing so, we highlighted the important discoveries that have been made via mathematical modeling and identified the gaps that still need to be filled. All the mathematical studies, reviewed here, indicate that combining cancer treatments strategies leads to enhanced therapeutic outcomes. Mathematical models of chemoimmunotherapy showed that with the use of this strategy for the treatment of chronic lymphocytic leukaemia, the patient's immune system would be boosted in addition to reducing the tumour volume. This appraoch leads to an increase in the number of T-cells that are activated when chemotherapy and immunotherapy are combined compared to a single therapy such as chemotherapy \cite{chappell2015mathematical}. Further, ligand transduced cells can trigger enough immune response to put tumour growth under control, whereas control-transduced tumour cells evade immunity \cite{de2009mathematical}. However, non of these models can yet provide an optimal dosage for the combination of chemotherapy and immunotherapy.

Chemovirotherapy has experienced notable successes both clinically and experimentally, however, much is still unknown about it \cite{malinzi2018enhancement}.  The principal mechanisms underlying the anti-tumoural effects of many ordinary virotherapies in conjunction with chemotherapeutics still remains a mystery. Fundamental questions such as prospective ramifications of chemotherapeutic agents on viral replication and the immunotherapeutic effects caused by virotherapy are yet to be fully answered. Another troubling problem is that of fully distinguishing chemotherapy-related toxicity from that of virotherapy induced toxicity \cite{binz2015chemovirotherapy}. Designing optimal scheduling of chemovirotherapy treatment attracts questions in an attempt to decipher the right dose combination, deciding the most effectual method of drug infusion and the critical treatment characteristics \cite{malinzi2018enhancement}. Furthermore, figuring out the potential implications of chemotherapeutics on viral reproduction and/or the immunotherapeutic effects possibly caused by virotherapeutics is never easy \cite{binz2015chemovirotherapy}. Combinations related to chemotherapy produce compounded toxicity and bone-marrow suppression and therefore their potential remains to be seen. Despite a great deal of clinical research in virotherapy, not many mathematical models of chemovirotherapy yet exist. Mathematical modeling has shown enhancement of chemotherapeutic drugs with oncolytic viruses leads to the depletion of all tumour cells from a patient's body tissue in less than a month \cite{malinzi2017modelling,malinzi2018enhancement,malinzi2019mathematical}. This time frame, nonetheless, leads to questions relating to toxicity. It is therefore important for future mathematical studies to investigate the effect of oncolytic viruses on the tumor microenvironment. Mathematical models of immunovirotherapy showed that the influx of immune cells around the tumour coupled with oncolytic viruses acts synergistically to lyse cancer cells. Nonetheless, one important aspect of oncolytic virotherapy that has not thoroughly been investigated, using mathematical models, is the effect of the viruses on normal body cells. Moreover, viruses can trigger anti-viral immune responses, and this as well needs further investigation. Other models have shown radiotherapy to be more effective when used synergistically with oncolytic virotherapy \cite{dingli2006mathematical,tao2007free}, including estimating optimal dosage combinations \cite{tao2007free}. Nonetheless, just like with other combinations, it is important that future studies focus on investigating the effect of using both treatments on body tissue. Mathematical models of chemoradiotherapy and radioimmunotherapy have shown that such combinations may lead to better treatment outcomes compared to a single approach. Whilst some studies \cite{goldie1988mathematical,2000single} have attempted to identify optimal dosage combinations, no mathematical study has investigated the toxicity of the combination despite the adverse effects of radiotherapy alone being well documented. Although radioimmunotherapy is a promising treatment strategy, questions regarding its long-term safety issues still exist. There still exist difficulties in estimating responses based on dosimetry and radiobiological models \cite{sharkey2011cancer}. An understanding of how these treatments can be best administered, how frequent the treatments can be administered and how to best integrate them with other agents to increase the overall response is yet to be developed \cite{sharkey2011cancer}. In spite of the fact that it is still possible to administer higher doses of radioactivity, it is still not known if the ratio of the tumour to that of non-tumour is higher. In addition, practical questions such as ascertaining the efficacy of the amalgamation specifically the abscopal effect, how to sequence the combination to give the best result and determining the optimal timing of the amalgamation are still open \cite{shun2019role}.

Several mathematical models reviewed in this article, for simplicity, ignore pertinent factors in cancer dynamics for example, the pharmacogenomics of cancer treatment and tumour-immune interactions yet these would ultimately influence the experimental outcomes and mathematical results. It is therefore essential for future studies to employ more complex mathematical models, considering detailed information for example the tumour microenvironment. Furthermore, future mathematical models should focus on using different tools other than differential equations to consider scenarios that cannot be captured by ODEs and PDEs. Generally, there is also a dire need for more research to focus on consolidating the discovered mathematical insights into clinical practice.

 In contrast to the combinations summarised in Table \ref{tab:comparisonsofthedifferenttherapies}, chemovirotherapy represents a rationally designed combination due to the improved overall survival benefits, nonoverlapping safety profiles, reduction of drug resistance, the impact on the tumour and of note is the preclinical evidence of synergy between the components. In support of this are the mathematical models on chemovirotherapy which have been presented in the previous sections. For example, a study by Malinzi \cite{malinzi2019mathematical} on mathematical analysis of chemovirotherapy indicated that the tumour is likely to be cleared from the body in less than a month if both chemotherapy and virotherapy are used. Further, the study contended that in the treatment of cancer, oncolytic viruses enhance chemotherapeutic drugs which is also in agreement with \cite{malinzi2018enhancement}. Nonetheless, designing an effective combo of viruses and chemotherapeutic drugs will need to have a perfect balance between immune anti-viral and anti-tumour responses.

Whilst some forms of cancer are incurable and treatment is only administered to improve the quality of life and/or prolong life, a combination therapy strategy that proves clinically to be of benefit to the patient taking into account the cost incurred will prove highly valuable. Today there are many treatment options for patients diagnosed with cancer than there were a decade or two ago. Some of these treatments can produce significant responses that could completely eradicate metastastatic tumours \cite{TreatmentResistance}. Nonetheless, these treatments all experience the same problem, that is, drug resistance. This resistance of cancer cells to drugs could ultimately lead to cancer relapse or recurrence. Molecular alterations may play a pivotal role in either acquired or intrinsic drug resistance and some of these alterations contributing to these resistance include mutation of the drug's molecular target, changes in the way the drug interacts with the tumour, broad cellular changes, and changes in the tumour microenvironment among others  \cite{TreatmentResistance}. Efforts are being made in order to understand how cells develop this resistance and more importantly, how it can be overcome. While everyone knows that finding a cure to cancer is a hard task, it is evident today that combining different strategies and minimising the chances that it recurs, is definitely the correct path to seek.

We anticipate our understanding of the mechanisms of cancer to ameliorate its burden in the near future, but the success will depend on a multidisciplinary approach across multiple sections of science, including mathematics, ecology, epidemiology, genetics and immunology. The early euphoria following breakthroughs exploiting combination therapeutic strategies has been tainted by the existence of numerous challenges. To this end, we highlight the outstanding open problems that still exist despite the reported success and numerous efforts in the research community to combat cancer using combination anti-cancer therapy.

Despite the growing expectations, in practice, combination cancer therapy strategies have demonstrated inconsistent results that can be attributed to the problems incurred when conjoining these molecularly targeted agents \cite{tolcher2018improving}. It is however hard to establish whether the heterogeneity of response can be ascribed to the complications of tumour biology, poor drug penetration into the tumour or suboptimal/uncoordinated exposure to the conjoined agents \cite{tolcher2018improving}. Furthermore, the problem of determining the ratio at which therapeutic agents should be mixed to produce maximal synergistic and minimal anti-cancer effects is unanswered. The problems of how to find the best drug combinations, as well as how to effectively combine therapy ex vivo in human cells to avoid doing thousands of clinical trials is still disturbing.

It will be interesting to know how to effectively combine therapeutic agents to produce effective results with appropriate sequencing of these agents. It is likely that combining two or more therapies will probably increase toxicity if not ideally designed and thus the challenge of optimally constructing these agents to circumvent the adverse side effects to a tolerable level is of great concern. This is likely to lead to suboptimal administration of drugs and thus encouraging drug resistance in the course of the treatment. However, in clinical trials, it has always been essential to de-escalate the dose until the drugs become least toxic, a process which is time-consuming and laborious.

In conclusion, while there has been some remarkable achievement and promise in the utilization of mathematical models in the combination treatment strategies of cancer, there is still a long way left to go. Since cancer is a complex group of diseases, their mechanisms of development are poorly understood and thus mathematical models developed, thus far, are relatively simple. Thus an ideal and novel combination therapy selected for cancer treatment should significantly lyse only cancer cells, increase the overall survival of a patient, reduce drug resistance, increase the efficacy and show preclinical evidence of synergy.

Mathematical modeling is making a significant contribution to the treatment of cancer. Thus given the pressure to find a value-based healthcare cancer therapy, there is absolutely no qualm that a mathematical approach will be an integral tool in determining the best combination strategy at a reduced cost while improving cancer treatment outcomes in the years to come. However, despite the promising results reported in various studies towards cancer treatment, most of them have not been integrated for routine clinical use. 

It is widely acknowledged that defeating cancer is unlikely to happen from a single therapeutic agent but rather a reasonable amalgamation of harmonious compatible strategies chosen from diverse therapeutics including traditional and modern cancer therapies. Ideally, there is a dire need to optimize dosing, duration as well as relative scheduling of the diverse components by considering the potential synergies and the probable antagonisms, as well as toxicities. With the numerous possibilities of combinatorial therapies and the outstanding challenge of figuring out the interplays between complex and intricated pharmacologic processes, we trust that optimization of these combinations in silico before testing them in patients could be made possible by developing and analysing integrated mathematical models.

Finally, we would like to note that this article does not provide an intricate review of the mathematical equations which constitute the models that have been discussed. Rather, a comprehensive review of mathematical models for combination cancer therapy has been carried out. We envisage that these models shall potentially play an essential role in deciding on strategies that are viable both medically and economically with a high likelihood of yielding clinical benefits.

\vspace*{2cm}

\section*{Additional information }

\subsection*{ Ethical approval and consent to participate }
This study is exempt from ethical approval in all the authors' affilitations.

\subsection*{ Consent to publish }
The authors guarantee that this work has not been previously published elsewhere.

\subsection*{ Data availability }
The authors declare that all data supporting the findings of this study are available within the article. Any other supplementary information that is not included in the article is available on request from the corresponding author. 

\subsection*{ Conflicts of interest }
The authors declare no conflict of interest. 

\subsection*{ Funding }
No author recieved funding to specifically carry out this study. 

\subsection*{Authors' contributions}
J.M conceptaulized the idea and designed the study. K.B.B and J.M wrote the initial draft of the article. H.A.A and S.P reviewed the manuscript and made critical inputs. All authors proof read and revised the article before submission. 

\subsection*{Acknowledgements}
H.A.A thanks the South African Medical Research Council (SAMRC) for a mid-career scientist and Self-initiated research grant; and the South African National Research Foundation (NRF) for Research Development Grant (RDG) for rated researchers.

\nocite{*}
\bibliographystyle{unsrt}
\bibliography{Main}

\begin{thebibliography}{100}

\bibitem{WHO2018}
{W}orld {H}ealth {O}rganisation.
\newblock \url{https://www.who.int/news-room/fact-sheets/detail/cancer}.
\newblock Accessed: April 2020.

\bibitem{hu2016recent}
Q.~Hu, W.~Sun, C.~Wang, and Z.~Gu.
\newblock Recent advances of cocktail chemotherapy by combination drug delivery
  systems.
\newblock {\em Advanced Drug Delivery Reviews}, 98:19--34, 2016.

\bibitem{beil2002sequencing}
D.R. Beil and L.M. Wein.
\newblock Sequencing surgery, radiotherapy and chemotherapy: Insights from a
  mathematical analysis.
\newblock {\em Breast Cancer Research and Treatment}, 74(3):279--286, 2002.

\bibitem{benzekry2017contributions}
S.~Benzekry.
\newblock {\em Contributions in mathematical oncology: When theory meets
  reality}.
\newblock PhD thesis, Universit{\'e} de Bordeaux, 2017.

\bibitem{heath2008nanotechnology}
J.R. Heath and M.E. Davis.
\newblock Nanotechnology and cancer.
\newblock {\em Annual Review of Medicine}, 59:251--265, 2008.

\bibitem{byrne2010dissecting}
H.M. Byrne.
\newblock Dissecting cancer through mathematics: from the cell to the animal
  model.
\newblock {\em Nature Reviews Cancer}, 10(3):221--230, 2010.

\bibitem{araujo2004history}
R.P. Araujo and D.L.S. McElwain.
\newblock A history of the study of solid tumour growth: the contribution of
  mathematical modelling.
\newblock {\em Bulletin of Mathematical Biology}, 66(5):1039--1091, 2004.

\bibitem{barbolosi2016computational}
D.~Barbolosi, J.~Ciccolini, F.~Lacarelle, B.and~Barl{\'e}si, and N.~Andr{\'e}.
\newblock Computational oncology-mathematical modelling of drug regimens for
  precision medicine.
\newblock {\em Nature Reviews Clinical Oncology}, 13(4):242, 2016.

\bibitem{kuang2016introduction}
Y.~Kuang, J.D. Nagy, and S.E. Eikenberry.
\newblock {\em Introduction to mathematical oncology}, volume~59.
\newblock CRC Press, 2016.

\bibitem{chappell2015mathematical}
M.~Chappell, V.~Chelliah, M.~Cherkaoui, G.~Derks, T.~Dumortier, N.~Evans,
  M.~Ferrarini, C.~Fornari, P.~Ghazal, M.L. Guerriero, et~al.
\newblock Mathematical modelling for combinations of immuno-oncology and
  anti-cancer therapies.
\newblock In {\em Quantitative Systems Pharmacology UK Meeting, Macclesfeld},
  2015.

\bibitem{malinzi2017modelling}
J.~Malinzi, A.~Eladdadi, and P.~Sibanda.
\newblock Modelling the spatiotemporal dynamics of chemovirotherapy cancer
  treatment.
\newblock {\em Journal of Biological Dynamics}, 11(1):244--274, 2017.

\bibitem{malinzi2018enhancement}
J.~Malinzi, R.~Ouifki, A.~Eladdadi, D.F.M. Torres, and K.A. White.
\newblock Enhancement of chemotherapy using oncolytic virotherapy: Mathematical
  and optimal control analysis.
\newblock {\em Mathematical Biosciences and Engineering}, 15(6):1435, 2018.

\bibitem{goldie1988mathematical}
J.H. Goldie, A.J. Coldman, V.~Ng, H.A. Hopkins, and W.B. Looney.
\newblock A mathematical and computer-based model of alternating chemotherapy
  and radiation therapy in experimental neoplasms1.
\newblock In {\em Treatment Modalities in Lung Cancer}, volume~41, pages
  11--20. Karger Publishers, 1988.

\bibitem{2000single}
J.A. O'Donoghue, G.~Sgouros, C.R. Divgi, and J.L. Humm.
\newblock Single-dose versus fractionated radioimmunotherapy: model comparisons
  for uniform tumor dosimetry.
\newblock {\em Journal of Nuclear Medicine}, 41(3):538--547, 2000.

\bibitem{kim2015quantitative}
P.S. Kim, J.J. Crivelli, I.~Choi, C.~Yun, and J.R. Wares.
\newblock Quantitative impact of immunomodulation versus oncolysis with
  cytokine-expressing virus therapeutics.
\newblock {\em Mathematical Biosciences \& Engineering}, 12(4):841--858, 2015.

\bibitem{friedman2018combination}
A.~Friedman and X.~Lai.
\newblock Combination therapy for cancer with oncolytic virus and checkpoint
  inhibitor: A mathematical model.
\newblock {\em PloS One}, 13(2):e0192449, 2018.

\bibitem{dingli2006mathematical}
D.~Dingli, M.D. Cascino, K.~Josi{\'c}, S.J. Russell, and {\v{Z}.}~Bajzer.
\newblock Mathematical modeling of cancer radiovirotherapy.
\newblock {\em Mathematical Biosciences}, 2006.

\bibitem{hsieh2018adjuvant}
M.C. Hsieh, and H.H.~Yu W.W.~Chang, C.Y. Lu, C.L. Chang, J.M. Chow, S.U. Chen,
  Y.~Cheng, and S.Y. Wu.
\newblock Adjuvant radiotherapy and chemotherapy improve survival in patients
  with pancreatic adenocarcinoma receiving surgery: adjuvant chemotherapy alone
  is insufficient in the era of intensity modulation radiation therapy.
\newblock {\em Cancer Medicine}, 7(6):2328--2338, 2018.

\bibitem{trimble1993neoadjuvant}
E.L. Trimble, R.S. Ungerleider, J.A. Abrams, R.S. Kaplan, E.G. Feigal, M.A.
  Smith, C.L. Carter, and M.A. Friedman.
\newblock Neoadjuvant therapy in cancer treatment.
\newblock {\em Cancer}, 72(S11):3515--3524, 1993.

\bibitem{ottolino2010intelligent}
K.~Ottolino-Perry, J.~Diallo, B.D. Lichty, J.~C Bell, and J.A. McCart.
\newblock Intelligent design: combination therapy with oncolytic viruses.
\newblock {\em Molecular Therapy}, 18(2):251--263, 2010.

\bibitem{chan2016chemotherapy}
W.~Chan, K.~Lam, W.~Siu, and K.~Yuen.
\newblock Chemotherapy at end-of-life: An integration of oncology and
  palliative team.
\newblock {\em Supportive Care in Cancer}, 24(3):1421--1427, 2016.

\bibitem{malinzi2019mathematical}
J.~Malinzi.
\newblock Mathematical analysis of a mathematical model of chemovirotherapy:
  Effect of drug infusion method.
\newblock {\em Computational and Mathematical Methods in Medicine}, 2019, 2019.

\bibitem{mokhtari2017combination}
R.B Mokhtari, T.S. Homayouni, N.~Baluch, E.~Morgatskaya, S.~Kumar, B.~Das, and
  H.~Yeger.
\newblock Combination therapy in combating cancer.
\newblock {\em Oncotarget}, 8(23):38022, 2017.

\bibitem{dry2016looking}
J.R. Dry, M.~Yang, and J.~Saez-Rodriguez.
\newblock Looking beyond the cancer cell for effective drug combinations.
\newblock {\em Genome Medicine}, 8(1):125, 2016.

\bibitem{doroshow2017design}
J.H. Doroshow and R.M. Simon.
\newblock On the design of combination cancer therapy.
\newblock {\em Cell}, 171(7):1476--1478, 2017.

\bibitem{lopez2017combine}
J.S. Lopez and U.~Banerji.
\newblock Combine and conquer: challenges for targeted therapy combinations in
  early phase trials.
\newblock {\em Nature Reviews Clinical Oncology}, 14(1):57, 2017.

\bibitem{emens2010chemoimmunotherapy}
L.A. Emens.
\newblock Chemoimmunotherapy.
\newblock {\em Cancer Journal (Sudbury, Mass.)}, 16(4):295, 2010.

\bibitem{de2009mathematical}
L.~de~Pillis, K.R. Fister, W.~Gu, C.~Collins, M.~Daub, D.~Gross, J.~Moore, and
  B.~Preskill.
\newblock Mathematical model creation for cancer chemo-immunotherapy.
\newblock {\em Computational and Mathematical Methods in Medicine},
  10(3):165--184, 2009.

\bibitem{de2007mathematical}
S.~De~Lillo, M.C. Salvatori, and N.~Bellomo.
\newblock Mathematical tools of the kinetic theory of active particles with
  some reasoning on the modelling progression and heterogeneity.
\newblock {\em Mathematical and Computer Modelling}, 45(5-6):564--578, 2007.

\bibitem{adam1997general}
J.A. Adam.
\newblock General aspects of modeling tumor growth and immune response.
\newblock In {\em A survey of Models for Tumor-Immune System Dynamics}, pages
  15--87. Springer, 1997.

\bibitem{chaplain2006mathematical}
M.~Chaplain and A.~Matzavinos.
\newblock Mathematical modelling of spatio-temporal phenomena in tumour
  immunology.
\newblock In {\em Tutorials in Mathematical Biosciences III}, pages 131--183.
  Springer, 2006.

\bibitem{de1985macrophage}
R.J. De~Boer, P.~Hogeweg, H.F Dullens, R.A. De~Weger, and W.~Den~Otter.
\newblock Macrophage {T} lymphocyte interactions in the anti-tumor immune
  response: a mathematical model.
\newblock {\em The Journal of Immunology}, 134(4):2748--2758, 1985.

\bibitem{arciero2004mathematical}
J.C. Arciero, T.L. Jackson, and D.E. Kirschner.
\newblock A mathematical model of tumor-immune evasion and sirna treatment.
\newblock {\em Discrete and Continuous Dynamical Systems Series B},
  4(1):39--58, 2004.

\bibitem{de2006mixed}
L.G. de~Pillis, W.~Gu, and A.E. Radunskaya.
\newblock Mixed immunotherapy and chemotherapy of tumors: modeling,
  applications and biological interpretations.
\newblock {\em Journal of Theoretical Biology}, 238(4):841--862, 2006.

\bibitem{de2005validated}
L.G. de~Pillis, A.E. Radunskaya, and C.L. Wiseman.
\newblock A validated mathematical model of cell-mediated immune response to
  tumor growth.
\newblock {\em Cancer Research}, 65(17):7950--7958, 2005.

\bibitem{isaeva2009different}
O.G. Isaeva and V.A. Osipov.
\newblock Different strategies for cancer treatment: mathematical modelling.
\newblock {\em Computational and Mathematical Methods in Medicine},
  10(4):253--272, 2009.

\bibitem{kirschner1998modeling}
D.~Kirschner and J.C. Panetta.
\newblock Modeling immunotherapy of the tumor--immune interaction.
\newblock {\em Journal of Mathematical Biology}, 37(3):235--252, 1998.

\bibitem{deng2014irradiation}
L.~Deng, H.~Liang, B.~Burnette, M.~Beckett, T.~Darga, R.R. Weichselbaum, and
  Y.~Fu.
\newblock Irradiation and anti--pd-l1 treatment synergistically promote
  antitumor immunity in mice.
\newblock {\em The Journal of Clinical Investigation}, 124(2):687--695, 2014.

\bibitem{de2003mathematical}
L.G. de~Pillis and A.~Radunskaya.
\newblock A mathematical model of immune response to tumor invasion.
\newblock In {\em Computational Fluid and Solid Mechanics 2003}, pages
  1661--1668. Elsevier, 2003.

\bibitem{rodrigues2019mathematical}
D.S. Rodrigues, P.F.A. Mancera, T.~Carvalho, and L.F. Gon{\c{c}}alves.
\newblock A mathematical model for chemoimmunotherapy of chronic lymphocytic
  leukemia.
\newblock {\em Applied Mathematics and Computation}, 349:118--133, 2019.

\bibitem{pang2016mathematical}
L.~Pang, L.~Shen, and Z.~Zhao.
\newblock Mathematical modelling and analysis of the tumor treatment regimens
  with pulsed immunotherapy and chemotherapy.
\newblock {\em Computational and Mathematical Methods in Medicine}, 2016, 2016.

\bibitem{vaha2007oncolytic}
J.V. V{\"a}h{\"a}-Koskela, J.E. Heikkil{\"a}, and A.E. Hinkkanen.
\newblock Oncolytic viruses in cancer therapy.
\newblock {\em Cancer Letters}, 254(2):178--216, 2007.

\bibitem{guo2008oncolytic}
Z.S. Guo, S.H. Thorne, and D.L. Bartlett.
\newblock Oncolytic virotherapy: molecular targets in tumor-selective
  replication and carrier cell-mediated delivery of oncolytic viruses.
\newblock {\em Biochimica et Biophysica Acta (BBA)-Reviews on Cancer},
  1785(2):217--231, 2008.

\bibitem{fukuhara2016oncolytic}
H.~Fukuhara, Y.~Ino, and T.~Todo.
\newblock Oncolytic virus therapy: A new era of cancer treatment at dawn.
\newblock {\em Cancer Science}, 107(10):1373--1379, 2016.

\bibitem{li2012oncolytic}
S.~Li, J.~Tong, M.M. Rahman, T.G. Shepherd, and G.~McFadden.
\newblock Oncolytic virotherapy for ovarian cancer.
\newblock {\em Oncolytic Virotherapy}, 1:1, 2012.

\bibitem{yu2009oncolytic}
Z.~Yu, S.~Li, P.~Brader, N.~Chen, A.Y. Yong, Q.~Zhang, A.A. Szalay, Y.~Fong,
  and R.J. Wong.
\newblock Oncolytic vaccinia therapy of squamous cell carcinoma.
\newblock {\em Molecular Cancer}, 8(1):45, 2009.

\bibitem{nguyen2014chemotherapy}
A.~Nguyen, L.~Ho, and Y.~Wan.
\newblock Chemotherapy and oncolytic virotherapy: advanced tactics in the war
  against cancer.
\newblock {\em Frontiers in Oncology}, 4:145, 2014.

\bibitem{ungerechts2010mantle}
G.~Ungerechts, M.E. Frenzke, K.~Yaiw, T.~Miest, P.B. Johnston, and R.~Cattaneo.
\newblock Mantle cell lymphoma salvage regimen: synergy between a reprogrammed
  oncolytic virus and two chemotherapeutics.
\newblock {\em Gene Therapy}, 17(12):1506, 2010.

\bibitem{ulasov2009combination}
I.V. Ulasov, A.M. Sonabend, S.~Nandi, A.~Khramtsov, Y.~Han, and M.S. Lesniak.
\newblock Combination of adenoviral virotherapy and temozolomide chemotherapy
  eradicates malignant glioma through autophagic and apoptotic cell death in
  vivo.
\newblock {\em British Journal of Cancer}, 100(7):1154, 2009.

\bibitem{friedman2006glioma}
A.~Friedman, J.P. Tian, G.~Fulci, E.A. Chiocca, and J.~Wang.
\newblock Glioma virotherapy: effects of innate immune suppression and
  increased viral replication capacity.
\newblock {\em Cancer Research}, 66(4):2314--2319, 2006.

\bibitem{friedman2003analysis}
A.~Friedman and Y.~Tao.
\newblock Analysis of a model of a virus that replicates selectively in tumor
  cells.
\newblock {\em Journal of Mathematical Biology}, 47(5):391--423, 2003.

\bibitem{wein2003validation}
L.M. Wein, J.T. Wu, and D.H. Kirn.
\newblock Validation and analysis of a mathematical model of a
  replication-competent oncolytic virus for cancer treatment: implications for
  virus design and delivery.
\newblock {\em Cancer Research}, 63(6):1317--1324, 2003.

\bibitem{wu2001modeling}
J.T. Wu, H.M. Byrne, D.H. Kirn, and L.M. Wein.
\newblock Modeling and analysis of a virus that replicates selectively in tumor
  cells.
\newblock {\em Bulletin of Mathematical Biology}, 63(4):731, 2001.

\bibitem{tao2005competitive}
Y.~Tao and Q.~Guo.
\newblock The competitive dynamics between tumor cells, a replication-competent
  virus and an immune response.
\newblock {\em Journal of Mathematical Biology}, 51(1):37--74, 2005.

\bibitem{eftimie2018tumour}
R.~Eftimie and G.~Eftimie.
\newblock Tumour-associated macrophages and oncolytic virotherapies: a
  mathematical investigation into a complex dynamics.
\newblock {\em Letters in Biomathematics}, 5(sup1):S6--S35, 2018.

\bibitem{JPH14}
J.~Malinzi, P.~Sibanda, and H.~Mambili-Mamoboundou.
\newblock Analysis of virotherapy in solid tumor invasion.
\newblock {\em Journal of Mathematical Biosciences}, 263:102--110, 2015.

\bibitem{steel1979exploitable}
G.G. Steel and Michael~J. Peckham.
\newblock Exploitable mechanisms in combined radiotherapy-chemotherapy: the
  concept of additivity.
\newblock {\em International Journal of Radiation Oncology Biology, Physics},
  5(1):85--91, 1979.

\bibitem{grassberger2016methodologies}
C.~Grassberger and H.~Paganetti.
\newblock Methodologies in the modeling of combined chemo-radiation treatments.
\newblock {\em Physics in Medicine \& Biology}, 61(21):R344, 2016.

\bibitem{ergun2003optimal}
A.~Ergun, K.~Camphausen, and L.M. Wein.
\newblock Optimal scheduling of radiotherapy and angiogenic inhibitors.
\newblock {\em Bulletin of Mathematical Biology}, 65(3):407--424, 2003.

\bibitem{salari2015mathematical}
E.~Salari, J.~Unkelbach, and T.~Bortfeld.
\newblock A mathematical programming approach to the fractionation problem in
  chemoradiotherapy.
\newblock {\em IIE Transactions on Healthcare Systems Engineering},
  5(2):55--73, 2015.

\bibitem{medina2012can}
J.~Medina-Echeverz and P.~Berraondo.
\newblock How can chemoimmunotherapy best be used for the treatment of colon
  cancer?
\newblock {\em Immunotherapy}, 4(12):1787--1790, 2012.

\bibitem{ghaffari2016mixed}
A.~Ghaffari, B.~Bahmaie, and M.~Nazari.
\newblock A mixed radiotherapy and chemotherapy model for treatment of cancer
  with metastasis.
\newblock {\em Mathematical Methods in the Applied Sciences},
  39(15):4603--4617, 2016.

\bibitem{marcu2006scheduling}
L.~Marcu, E.~Bezak, and I.~Olver.
\newblock Scheduling cisplatin and radiotherapy in the treatment of squamous
  cell carcinomas of the head and neck: a modelling approach.
\newblock {\em Physics in Medicine \& Biology}, 51(15):3625, 2006.

\bibitem{barazzuol2010mathematical}
L.~Barazzuol, N.G. Burnet, R.~Jena, B.~Jones, S.J. Jefferies, and N.F. Kirkby.
\newblock A mathematical model of brain tumour response to radiotherapy and
  chemotherapy considering radiobiological aspects.
\newblock {\em Journal of Theoretical Biology}, 262(3):553--565, 2010.

\bibitem{powathil2012modelling}
G.G. Powathil, K.E. Gordon, L.A. Hill, and M.A.J. Chaplain.
\newblock Modelling the effects of cell-cycle heterogeneity on the response of
  a solid tumour to chemotherapy: biological insights from a hybrid multiscale
  cellular automaton model.
\newblock {\em Journal of Theoretical Biology}, 308:1--19, 2012.

\bibitem{jones1999inclusion}
B.~Jones and R.G. Dale.
\newblock Inclusion of molecular biotherapies with radical radiotherapy:
  modeling of combined modality treatment schedules.
\newblock {\em International Journal of Radiation Oncology, Biology and
  Physics}, 45(4):1025--1034, 1999.

\bibitem{jones2005potential}
B.~Jones and R.G. Dale.
\newblock The potential for mathematical modelling in the assessment of the
  radiation dose equivalent of cytotoxic chemotherapy given concomitantly with
  radiotherapy.
\newblock {\em The British Journal of Radiology}, 78(934):939--944, 2005.

\bibitem{plataniotis2008use}
G.A. Plataniotis and R.G. Dale.
\newblock Use of concept of chemotherapy-equivalent biologically effective dose
  to provide quantitative evaluation of contribution of chemotherapy to local
  tumor control in chemoradiotherapy cervical cancer trials.
\newblock {\em International Journal of Radiation Oncology, Biology and
  Physics}, 72(5):1538--1543, 2008.

\bibitem{plataniotis2014assessment}
G.A. Plataniotis and R.G. Dale.
\newblock Assessment of the radiation-equivalent of chemotherapy contributions
  in 1-phase radio-chemotherapy treatment of muscle-invasive bladder cancer.
\newblock {\em International Journal of Radiation Oncology, Biology and
  Physics}, 88(4):927--932, 2014.

\bibitem{moraru2014radiation}
I.C. Moraru, A.~Tai, B.~Erickson, and X.A. Li.
\newblock Radiation dose responses for chemoradiation therapy of pancreatic
  cancer: an analysis of compiled clinical data using biophysical models.
\newblock {\em Practical Radiation Oncology}, 4(1):13--19, 2014.

\bibitem{durante2015modeling}
M.~Durante, F.~Tommasino, and S.~Yamada.
\newblock Modeling combined chemotherapy and particle therapy for locally
  advanced pancreatic cancer.
\newblock {\em Frontiers in Oncology}, 5:145, 2015.

\bibitem{hartley2010radiobiological}
A.~Hartley, P.~Sanghera, J.~Glaholm, H.~Mehanna, C.~McConkey, and J.~Fowler.
\newblock Radiobiological modelling of the therapeutic ratio for the addition
  of synchronous chemotherapy to radiotherapy in locally advanced squamous cell
  carcinoma of the head and neck.
\newblock {\em Clinical Oncology}, 22(2):125--130, 2010.

\bibitem{meade2013revised}
S.~Meade, C.~McConkey, P.~Sanghera, H.~Mehanna, and A.~Hartley.
\newblock Revised radiobiological modelling of the contribution of synchronous
  chemotherapy to the rate of grades 3--4 mucositis in head and neck cancer.
\newblock {\em Journal of Medical Imaging and Radiation Oncology},
  57(6):733--738, 2013.

\bibitem{meade2013revising}
S.~Meade, P.~Sanghera, C.~McConkey, J.~Fowler, G.~Fountzilas, J.~Glaholm, and
  A.~Hartley.
\newblock Revising the radiobiological model of synchronous chemotherapy in
  head-and-neck cancer: a new analysis examining reduced weighting of
  accelerated repopulation.
\newblock {\em International Journal of Radiation Oncology, Biology and
  Physics}, 86(1):157--163, 2013.

\bibitem{sgouros1992plasmapheresis}
G.~Sgouros.
\newblock Plasmapheresis in radioimmunotherapy of micrometastases: a
  mathematical modeling and dosimetrical analysis.
\newblock {\em Journal of Nuclear Medicine: official publication, Society of
  Nuclear Medicine}, 33(12):2167--2179, 1992.

\bibitem{epenetos1986limitations}
A.A. Epenetos, D.~Snook, H.~Durbin, P.M. Johnson, and J~Taylor-Papadimitriou.
\newblock Limitations of radiolabeled monoclonal antibodies for localization of
  human neoplasms.
\newblock {\em Cancer Research}, 46(6):3183--3191, 1986.

\bibitem{order1990radiolabeled}
S.E. Order, A.M. Sleeper, G.B. Stillwagon, J.L. Klein, and P.K. Leichner.
\newblock Radiolabeled antibodies: results and potential in cancer therapy.
\newblock {\em Cancer Research}, 50(3 Supplement):1011s--1013s, 1990.

\bibitem{begent1990antibody}
R.H.J. Begent and R.B. Pedley.
\newblock Antibody targeted therapy in cancer: comparison of murine and
  clinical studies.
\newblock {\em Cancer Treatment Reviews}, 17(2-3):373--378, 1990.

\bibitem{sharkey1990biological}
R.M. Sharkey, R.D. Blumenthal, H.J. Hansen, and D.M. Goldenberg.
\newblock Biological considerations for radioimmunotherapy.
\newblock {\em Cancer Research}, 50(3 Supplement):964s--969s, 1990.

\bibitem{fujimori1989modeling}
K.~Fujimori, D.G. Covell, J.E. Fletcher, and J.N. Weinstein.
\newblock Modeling analysis of the global and microscopic distribution of
  immunoglobulin {G}, {F} (ab')2, and fab in tumors.
\newblock {\em Cancer Research}, 49(20):5656--5663, 1989.

\bibitem{fujimori1990modeling}
K.~Fujimori, D.G. Covell, J.E. Fletcher, and J.N. Weinstein.
\newblock A modeling analysis of monoclonal antibody percolation through
  tumors: a binding-site barrier.
\newblock {\em Journal of Nuclear Medicine: Official Publication, Society of
  Nuclear Medicine}, 31(7):1191--1198, 1990.

\bibitem{fujimori1991integrated}
K.~Fujimori, D.R. Fisher, and J.N. Weinstein.
\newblock Integrated microscopic-macroscopic pharmacology of monoclonal
  antibody radioconjugates: the radiation dose distribution.
\newblock {\em Cancer Research}, 51(18):4821--4827, 1991.

\bibitem{van1991analysis}
W.~van Osdol, K.~Fujimori, and J.N. Weinstein.
\newblock An analysis of monoclonal antibody distribution in microscopic tumor
  nodules: consequences of a ``binding site barrier".
\newblock {\em Cancer Research}, 51(18):4776--4784, 1991.

\bibitem{kumar2010mathematical}
D.~Kumar and S.~Kumar.
\newblock A mathematical model of radioimmunotherapy for tumor treatment.
\newblock {\em African Journal of Mathematics and Computer Science Research},
  3(6):101--106, 2010.

\bibitem{flux1997three}
G.D. Flux, S.~Webb, R.J. Ott, S.J. Chittenden, and R.~Thomas.
\newblock Three-dimensional dosimetry for intralesional radionuclide therapy
  using mathematical modeling and multimodality imaging.
\newblock {\em Journal of Nuclear Medicine}, 38(7):1059--1066, 1997.

\bibitem{serre2016mathematical}
R.~Serre, S.~Benzekry, L.~Padovani, C.~Meille, N.~Andr{\'e}, J.~Ciccolini,
  F.~Barlesi, X.~Muracciole, and D.~Barbolosi.
\newblock Mathematical modeling of cancer immunotherapy and its synergy with
  radiotherapy.
\newblock {\em Cancer Research}, 76(17):4931--4940, 2016.

\bibitem{melcher2011thunder}
A.~Melcher, K.~Parato, C.M. Rooney, and J.C. Bell.
\newblock Thunder and lightning: immunotherapy and oncolytic viruses collide.
\newblock {\em Molecular Therapy}, 19(6):1008--1016, 2011.

\bibitem{cheema2014immunovirotherapy}
T.A. Cheema, P.E. Fecci, J.~Ning, and S.D. Rabkin.
\newblock Immunovirotherapy for the treatment of glioblastoma.
\newblock {\em Oncoimmunology}, 3(1):e27218, 2014.

\bibitem{hardcastle2017immunovirotherapy}
J.~Hardcastle, L.~Mills, C.S. Malo, F.~Jin, C.~Kurokawa, H.~Geekiyanage,
  M.~Schroeder, J.~Sarkaria, A.J. Johnson, and E.~Galanis.
\newblock Immunovirotherapy with measles virus strains in combination with
  anti--pd-1 antibody blockade enhances antitumor activity in glioblastoma
  treatment.
\newblock {\em Neuro-oncology}, 19(4):493--502, 2017.

\bibitem{cerullo2012oncolytic}
V.~Cerullo, M.~V{\"a}h{\"a}-Koskela, and A.~Hemminki.
\newblock Oncolytic adenoviruses: A potent form of tumor immunovirotherapy.
\newblock {\em Oncoimmunology}, 1(6):979--981, 2012.

\bibitem{wares2015treatment}
J.~R. Wares, J.J. Crivelli, C.~Yun, I.~Choi, J.L. Gevertz, and P.S. Kim.
\newblock Treatment strategies for combining immunostimulatory oncolytic virus
  therapeutics with dendritic cell injections.
\newblock {\em Mathematical Biosciences \& Engineering}, 12(6):1237--1256,
  2015.

\bibitem{bagheri2011dynamical}
N.~Bagheri, M.~Shiina, D.A. Lauffenburger, and W.M. Korn.
\newblock A dynamical systems model for combinatorial cancer therapy enhances
  oncolytic adenovirus efficacy by mek-inhibition.
\newblock {\em PLoS Computational Biology}, 7(2):e1001085, 2011.

\bibitem{huang2010therapeutic}
J.~Huang, S.~Zhang, K.~Choi, I.~Choi, J.~Kim, M.~Lee, H.~Kim, and C.~Yun.
\newblock Therapeutic and tumor-specific immunity induced by combination of
  dendritic cells and oncolytic adenovirus expressing il-12 and 4-1bbl.
\newblock {\em Molecular Therapy}, 18(2):264--274, 2010.

\bibitem{dingli2009dynamics}
D.~Dingli, C.~Offord, R.~Myers, K.~Peng, T.W. Carr, K.~Josic, S.J. Russell, and
  Z.~Bajzer.
\newblock Dynamics of multiple myeloma tumor therapy with a recombinant measles
  virus.
\newblock {\em Cancer Gene Therapy}, 16(12):873, 2009.

\bibitem{rommelfanger2012dynamics}
D.M. Rommelfanger, C.P. Offord, J.~Dev, Z.~Bajzer, R.G. Vile, and D.~Dingli.
\newblock Dynamics of melanoma tumor therapy with vesicular stomatitis virus:
  explaining the variability in outcomes using mathematical modeling.
\newblock {\em Gene Therapy}, 19(5):543, 2012.

\bibitem{mok2009mathematical}
W.~Mok, T.~Stylianopoulos, Y.~Boucher, and R.K. Jain.
\newblock Mathematical modeling of herpes simplex virus distribution in solid
  tumors: implications for cancer gene therapy.
\newblock {\em Clinical Cancer Research}, 15(7):2352--2360, 2009.

\bibitem{depillis2013model}
L.~DePillis, A.~Gallegos, and A.~Radunskaya.
\newblock A model of dendritic cell therapy for melanoma.
\newblock {\em Frontiers in Oncology}, 3, 2013.

\bibitem{pappalardo2014induction}
F.~Pappalardo, M.~Pennisi, A.~Ricupito, F.~Topputo, and M.~Bellone.
\newblock Induction of t-cell memory by a dendritic cell vaccine: a
  computational model.
\newblock {\em Bioinformatics}, 30(13):1884--1891, 2014.

\bibitem{dritschel2018mathematical}
H.~Dritschel, S.L. Waters, A.~Roller, and H.M. Byrne.
\newblock A mathematical model of cytotoxic and helper {T} cell interactions in
  a tumour microenvironment.
\newblock {\em Letters in Biomathematics}, 5(sup1):S36--S68, 2018.

\bibitem{wu2004analysis}
J.T. Wu, D.H. Kirn, and L.M. Wein.
\newblock Analysis of a three-way race between tumor growth, a
  replication-competent virus and an immune response.
\newblock {\em Bulletin of Mathematical Biology}, 66(4):605, 2004.

\bibitem{berg2019vitro}
D.R. Berg, C.P. Offord, I.~Kemler, M.K. Ennis, L.~Chang, G.~Paulik, Z.~Bajzer,
  C.~Neuhauser, and D.~Dingli.
\newblock In vitro and in silico multidimensional modeling of oncolytic tumor
  virotherapy dynamics.
\newblock {\em PLoS Computational Biology}, 15(3):e1006773, 2019.

\bibitem{wodarz2003gene}
D.~Wodarz.
\newblock Gene therapy for killing p53-negative cancer cells: use of
  replicating versus nonreplicating agents.
\newblock {\em Human Gene Therapy}, 14(2):153--159, 2003.

\bibitem{wodarz2012complex}
D~Wodarz, A.~Hofacre, J.W. Lau, Z.~Sun, H.~Fan, and N.L. Komarova.
\newblock Complex spatial dynamics of oncolytic viruses in vitro: mathematical
  and experimental approaches.
\newblock {\em PLoS Computational Biology}, 8(6):e1002547, 2012.

\bibitem{bajzer2008modeling}
{\v{Z}.}~Bajzer, T.~Carr, K.~Josi{\'c}, S.J. Russell, and D.~Dingli.
\newblock Modeling of cancer virotherapy with recombinant measles viruses.
\newblock {\em Journal of Theoretical Biology}, 252(1):109--122, 2008.

\bibitem{biesecker2010optimization}
M.~Biesecker, J~K., H.~Lu, D.~Dingli, and {\v{Z}}.~Bajzer.
\newblock Optimization of virotherapy for cancer.
\newblock {\em Bulletin of Mathematical Biology}, 72(2):469--489, 2010.

\bibitem{gevertz2018developing}
J.L. Gevertz and J.R. Wares.
\newblock Developing a minimally structured mathematical model of cancer
  treatment with oncolytic viruses and dendritic cell injections.
\newblock {\em Computational and Mathematical Methods in Medicine}, 2018, 2018.

\bibitem{timalsina2017mathematical}
A.~Timalsina, J.P. Tian, and J.~Wang.
\newblock Mathematical and computational modeling for tumor virotherapy with
  mediated immunity.
\newblock {\em Bulletin of Mathematical Biology}, 79(8):1736--1758, 2017.

\bibitem{jenner2018modelling}
A.L. Jenner, C.~Yun, A.~Yoon, A.C.F. Coster, and P.S. Kim.
\newblock Modelling combined virotherapy and immunotherapy: strengthening the
  antitumour immune response mediated by il-12 and gm-csf expression.
\newblock {\em Letters in Biomathematics}, 5(sup1):S99--S116, 2018.

\bibitem{mahasa2017oncolytic}
K.~J. Mahasa, A.~Eladdadi, L.~De~Pillis, and R.~Ouifki.
\newblock Oncolytic potency and reduced virus tumor-specificity in oncolytic
  virotherapy. a mathematical modelling approach.
\newblock {\em PloS One}, 12(9):e0184347, 2017.

\bibitem{cassidy2018mathematical}
T.~Cassidy and A.R. Humphries.
\newblock {A mathematical model of viral oncology as an immuno-oncology
  instigator}.
\newblock {\em Mathematical Medicine and Biology: A Journal of the IMA}, 04
  2019.

\bibitem{wodarz2001viruses}
D.~Wodarz.
\newblock Viruses as antitumor weapons: defining conditions for tumor
  remission.
\newblock {\em Cancer Research}, 61(8):3501--3507, 2001.

\bibitem{tao2007free}
Y.~Tao and Q.~Guo.
\newblock A free boundary problem modelling cancer radiovirotherapy.
\newblock {\em Mathematical Models and Methods in Applied Sciences},
  17(08):1241--1259, 2007.

\bibitem{NCI}
{N}ational {C}ancer {I}nstitute.
\newblock
  \url{https://www.cancer.gov/about-cancer/treatment/types/targeted-therapies/targeted-therapies-fact-sheet}.
\newblock Accessed: October 2020.

\bibitem{abbott2006mathematical}
L.H. Abbott and F.~Michor.
\newblock Mathematical models of targeted cancer therapy.
\newblock {\em British Journal of Cancer}, 95(9):1136--1141, 2006.

\bibitem{bozic2013evolutionary}
I.~Bozic, J.G. Reiter, B.~Allen, T.~Antal, K.~Chatterjee, P.~Shah, Y.~Moon,
  A.~Yaqubie, N.~Kelly, D.T. Le, et~al.
\newblock Evolutionary dynamics of cancer in response to targeted combination
  therapy.
\newblock {\em eLife}, 2:e00747, 2013.

\bibitem{yu2019combination}
C.~Yu, X.~Liu, J.~Yang, M.~Zhang, H.~Jin, X.~Ma, and H.~Shi.
\newblock Combination of immunotherapy with targeted therapy: Theory and
  practice in metastatic melanoma.
\newblock {\em Frontiers in Immunology}, 10:990, 2019.

\bibitem{green2001mathematical}
A.J. Green, C.J. Johnson, K.L. Adamson, and R.H.J. Begent.
\newblock Mathematical model of antibody targeting: important parameters
  defined using clinical data.
\newblock {\em Physics in Medicine \& Biology}, 46(6):1679, 2001.

\bibitem{shen2020biphasic}
J.~Shen, L.~Li, T.~Yang, P.S. Cohen, and G.~Sun.
\newblock Biphasic mathematical model of cell--drug interaction that separates
  target-specific and off-target inhibition and suggests potent targeted drug
  combinations for multi-driver colorectal cancer cells.
\newblock {\em Cancers}, 12(2):436, 2020.

\bibitem{sun2016mathematical}
X.~Sun, J.~guang Bao, and Y.~Shao.
\newblock Mathematical modeling of therapy-induced cancer drug resistance:
  connecting cancer mechanisms to population survival rates.
\newblock {\em Scientific Reports}, 6:22498, 2016.

\bibitem{kozlowska2018mathematical}
E.~Koz{\l}owska, A.~F{\"a}rkkil{\"a}, T.~Vallius, O.~Carp{\'e}n, J.~Kemppainen,
  S.~Gr{\'e}nman, R.~Lehtonen, J.~Hynninen, S.~Hietanen, and S.~Hautaniemi.
\newblock Mathematical modeling predicts response to chemotherapy and drug
  combinations in ovarian cancer.
\newblock {\em Cancer Research}, 78(14):4036--4044, 2018.

\bibitem{jarrett2019experimentally}
A.M. Jarrett, A.~Shah, M.J. Bloom, M.T McKenna, D.A. Hormuth, T.E. Yankeelov,
  and Anna~G. Sorace.
\newblock Experimentally-driven mathematical modeling to improve combination
  targeted and cytotoxic therapy for {HER2+} breast cancer.
\newblock {\em Scientific reports}, 9(1):1--12, 2019.

\bibitem{owen2011mathematical}
M.R. Owen, I.J. Stamper, M.~Muthana, G.W. Richardson, J.~Dobson, C.E. Lewis,
  and H.M. Byrne.
\newblock Mathematical modeling predicts synergistic antitumor effects of
  combining a macrophage-based, hypoxia-targeted gene therapy with
  chemotherapy.
\newblock {\em Cancer Research}, 71(8):2826--2837, 2011.

\bibitem{tang2013target}
J.~Tang, L.~Karhinen, T.~Xu, A.~Szwajda, B.~Yadav, K.~Wennerberg, and
  T.~Aittokallio.
\newblock Target inhibition networks: predicting selective combinations of
  druggable targets to block cancer survival pathways.
\newblock {\em PLoS Computational Biology}, 9(9):e1003226, 2013.

\bibitem{tang2019network}
J.~Tang, P.~Gautam, A.~Gupta, L.~He, S.~Timonen, Y.~Akimov, W.~Wang,
  A.~Szwajda, A.~Jaiswal, D.~Turei, et~al.
\newblock Network pharmacology modeling identifies synergistic aurora {B} and
  {ZAK} interaction in triple-negative breast cancer.
\newblock {\em {NPJ} Systems Biology and Applications}, 5(1):1--11, 2019.

\bibitem{araujo2005mathematical}
R.P. Araujo, E.F. Petricoin, and L.A. Liotta.
\newblock A mathematical model of combination therapy using the egfr signaling
  network.
\newblock {\em Biosystems}, 80(1):57--69, 2005.

\bibitem{komarova2005drug}
N.L. Komarova and D.~Wodarz.
\newblock Drug resistance in cancer: principles of emergence and prevention.
\newblock {\em Proceedings of the National Academy of Sciences},
  102(27):9714--9719, 2005.

\bibitem{charusanti2004mathematical}
P.~Charusanti, X.~Hu, L.~Chen, D.~Neuhauser, and J.J. DiStefano~III.
\newblock A mathematical model of bcr-abl autophosphorylation, signaling
  through the crkl pathway, and gleevec dynamics in chronic myeloid leukemia.
\newblock {\em Discrete \& Continuous Dynamical Systems-B}, 4(1):99, 2004.

\bibitem{shen2002model}
S.~Shen, J.~Duan, R.F. Meredith, D.J. Buchsbaum, I.A. Brezovich, P.N. Pareek,
  and J.A. Bonner.
\newblock Model prediction of treatment planning for dose-fractionated
  radioimmunotherapy.
\newblock {\em Cancer: Interdisciplinary International Journal of the American
  Cancer Society}, 94(S4):1264--1269, 2002.

\bibitem{callahan2013two}
M.K. Callahan.
\newblock Two drugs are better than one-modeling drug combinations in cancer
  therapy.
\newblock {\em Science Translational Medicine}, 5(194):194ec116--194ec116,
  2013.

\bibitem{chakwizira2018mathematical}
A.~Chakwizira, J.~Ahlstedt, H.~Nittby~Redebrandt, and C.~Ceberg.
\newblock Mathematical modelling of the synergistic combination of radiotherapy
  and indoleamine-2, 3-dioxygenase (ido) inhibitory immunotherapy against
  glioblastoma.
\newblock {\em The British Journal of Radiology}, 91(1087):20170857, 2018.

\bibitem{monjazeb2013combined}
A.M. Monjazeb, S.K. Grossenbacher, G.D. Sckisel, R.~Canter, E.E. Sparger,
  W.~Culp, M.S. Kent, and W.J. Murphy.
\newblock Combined radiotherapy and immunotherapy using {CPG}
  oligodeoxynucleotides and indolamine 2, 3 dioxygenase (ido) blockade.
\newblock {\em Journal for Immunotherapy of Cancer}, 1(1):P256, 2013.

\bibitem{radunskaya2018mathematical}
A.~Radunskaya, R.~Kim, I.I. Woods, et~al.
\newblock Mathematical modeling of tumor immune interactions: a closer look at
  the role of a {PD-L1} inhibitor in cancer immunotherapy.
\newblock {\em Spora: A Journal of Biomathematics}, 4(1):25--41, 2018.

\bibitem{nazari2018mathematical}
F.~Nazari, A.T. Pearson, J.E. N{\"o}r, and T.L. Jackson.
\newblock A mathematical model for il-6-mediated, stem cell driven tumor growth
  and targeted treatment.
\newblock {\em PLoS Computational Biology}, 14(1):e1005920, 2018.

\bibitem{wang2016optimal}
S.~Wang and H.~Sch{\"a}ttler.
\newblock Optimal control of a mathematical model for cancer chemotherapy under
  tumor heterogeneity.
\newblock {\em Math. Biosci. and Engr.-MBE}, 13:1223--1240, 2016.

\bibitem{malinzi2015mathematical}
J.~Malinzi.
\newblock {\em Mathematical modeling of cancer treatments and the role of the
  immune system response to tumor invasion.}
\newblock PhD thesis, 2015.

\bibitem{spring2019illuminating}
B.Q. Spring, R.T. Lang, E..M Kercher, I.~Rizvi, R..M Wenham, J.R.
  Conejo-Garcia, T.~Hasan, R.A. Gatenby, and H.~Enderling.
\newblock Illuminating the numbers: Integrating mathematical models to optimize
  photomedicine dosimetry and combination therapies.
\newblock {\em Frontiers in Physics}, 7:46, 2019.

\bibitem{rihan2014delay}
F.~A. Rihan, DH~Abdelrahman, F.~Al-Maskari, F.~Ibrahim, and M.~A. Abdeen.
\newblock Delay differential model for tumour-immune response with
  chemoimmunotherapy and optimal control.
\newblock {\em Computational and Mathematical Methods in Medicine}, 2014, 2014.

\bibitem{oke2018optimal}
S.~Oke, M.~Matadi, and S.~Xulu.
\newblock Optimal control analysis of a mathematical model for breast cancer.
\newblock {\em Mathematical and Computational Applications}, 23(2):21, 2018.

\bibitem{mamat2013mathematical}
M.~Mamat, K.A. Subiyanto, and A.~Kartono.
\newblock Mathematical model of cancer treatments using immunotherapy,
  chemotherapy and biochemotherapy.
\newblock {\em Applied Mathematical Sciences}, 7(5):247--261, 2013.

\bibitem{lai2018modeling}
X.~Lai, A.~Stiff, M.~Duggan, R.~Wesolowski, W.E. Carson, and A.~Friedman.
\newblock Modeling combination therapy for breast cancer with bet and immune
  checkpoint inhibitors.
\newblock {\em Proceedings of the National Academy of Sciences},
  115(21):5534--5539, 2018.

\bibitem{joshi2009immunotherapies}
B.~Joshi, X.~Wang, S.~Banerjee, H.~Tian, A.~Matzavinos, and M.A.J. Chaplain.
\newblock On immunotherapies and cancer vaccination protocols: a mathematical
  modelling approach.
\newblock {\em Journal of Theoretical Biology}, 259(4):820--827, 2009.

\bibitem{kosinsky2018radiation}
Y.~Kosinsky, S.J. Dovedi, K.~Peskov, V.~Voronova, L.~Chu, H.~Tomkinson,
  N.~Al-Huniti, D.R. Stanski, and G.~Helmlinger.
\newblock Radiation and pd-(l) 1 treatment combinations: immune response and
  dose optimization via a predictive systems model.
\newblock {\em Journal for Immunotherapy of Cancer}, 6(1):17, 2018.

\bibitem{rihan2019optimal}
F.A. Rihan, S.~Lakshmanan, and H.~Maurer.
\newblock Optimal control of tumour-immune model with time-delay and
  immuno-chemotherapy.
\newblock {\em Applied Mathematics and Computation}, 353:147--165, 2019.

\bibitem{akman2018optimal}
T.~Akman~Y{\i}ld{\i}z, S.~Arshad, and D.~Baleanu.
\newblock Optimal chemotherapy and immunotherapy schedules for a cancer-obesity
  model with caputo time fractional derivative.
\newblock {\em Mathematical Methods in the Applied Sciences},
  41(18):9390--9407, 2018.

\bibitem{ratajczyk2018optimal}
E.~Ratajczyk, U.~Ledzewicz, and H.~Sch{\"a}ttler.
\newblock Optimal control for a mathematical model of glioma treatment with
  oncolytic therapy and {T}{N}{F}-$\alpha$ inhibitors.
\newblock {\em Journal of Optimization Theory and Applications},
  176(2):456--477, 2018.

\bibitem{sharma2016analysis}
S.~Sharma and G.P. Samanta.
\newblock Analysis of the dynamics of a tumor--immune system with chemotherapy
  and immunotherapy and quadratic optimal control.
\newblock {\em Differential Equations and Dynamical Systems}, 24(2):149--171,
  2016.

\bibitem{ledzewicz2014optimal}
U.~Ledzewicz and H.~Sch{\"a}ttler.
\newblock An optimal control approach to cancer chemotherapy with tumor--immune
  system interactions.
\newblock In {\em Mathematical Models of Tumor-Immune System Dynamics}, pages
  157--196. Springer, 2014.

\bibitem{nazari2015finite}
M.~Nazari, A.~Ghaffari, and F.~Arab.
\newblock Finite duration treatment of cancer by using vaccine therapy and
  optimal chemotherapy: state-dependent riccati equation control and extended
  kalman filter.
\newblock {\em Journal of Biological Systems}, 23(01):1--29, 2015.

\bibitem{parra2013mathematical}
Z.P. Parra-Guillen, P.~Berraondo, E.~Grenier, B.~Ribba, and I.~F. Troconiz.
\newblock Mathematical model approach to describe tumour response in mice after
  vaccine administration and its applicability to immune-stimulatory
  cytokine-based strategies.
\newblock {\em The AAPS Journal}, 15(3):797--807, 2013.

\bibitem{rodrigues2013mathematical}
D.S. Rodrigues, P.F. de~Arruda~Mancera, et~al.
\newblock Mathematical analysis and simulations involving chemotherapy and
  surgery on large human tumours under a suitable cell-kill functional
  response.
\newblock {\em Mathematical Biosciences \& Engineering}, 10(1):221--34, 2013.

\bibitem{villasana2010modeling}
M.~Villasana, G.~Ochoa, and S.~Aguilar.
\newblock Modeling and optimization of combined cytostatic and cytotoxic cancer
  chemotherapy.
\newblock {\em Artificial Intelligence in Medicine}, 50(3):163--173, 2010.

\bibitem{chareyron2009mixed}
S.~Chareyron and M.~Alamir.
\newblock Mixed immunotherapy and chemotherapy of tumors: feedback design and
  model updating schemes.
\newblock {\em Journal of Theoretical Biology}, 258(3):444--454, 2009.

\bibitem{ledzewicz2008optimal}
U.~Ledzewicz, H.~Schattler, and A.~d'Onofrio.
\newblock Optimal control for combination therapy in cancer.
\newblock In {\em 2008 47th IEEE Conference on Decision and Control}, pages
  1537--1542. IEEE, 2008.

\bibitem{d2009optimal}
A.~d'Onofrio, U.~Ledzewicz, H.~Maurer, and H.~Sch{\"a}ttler.
\newblock On optimal delivery of combination therapy for tumors.
\newblock {\em Mathematical Biosciences}, 222(1):13--26, 2009.

\bibitem{powathil2007mathematical}
G.~Powathil, M.~Kohandel, S.~Sivaloganathan, A.~Oza, and M.~Milosevic.
\newblock Mathematical modeling of brain tumors: effects of radiotherapy and
  chemotherapy.
\newblock {\em Physics in Medicine \& Biology}, 52(11):3291, 2007.

\bibitem{imbs2018revisiting}
D.~Imbs, R.~El~Cheikh, A.~Boyer, J.~Ciccolini, C.~Mascaux, B.~Lacarelle,
  F.~Barlesi, D.~Barbolosi, and S.~Benzekry.
\newblock Revisiting bevacizumab+ cytotoxics scheduling using mathematical
  modeling: Proof of concept study in experimental non-small cell lung
  carcinoma.
\newblock {\em CPT: Pharmacometrics \& Systems Pharmacology}, 7(1):42--50,
  2018.

\bibitem{tao2008mathematical}
Y.~Tao and Q.~Guo.
\newblock A mathematical model of combined therapies against cancer using
  viruses and inhibitors.
\newblock {\em Science in China Series A: Mathematics}, 51(12):2315--2329,
  2008.

\bibitem{hadjiandreou2014mathematical}
M.M. Hadjiandreou and G.D. Mitsis.
\newblock Mathematical modeling of tumor growth, drug-resistance, toxicity, and
  optimal therapy design.
\newblock {\em IEEE Transactions on Biomedical Engineering}, 61(2):415--425,
  2014.

\bibitem{su2016optimal}
Y.~Su, C.~Jia, and Y.~Chen.
\newblock Optimal control model of tumor treatment with oncolytic virus and mek
  inhibitor.
\newblock {\em BioMed Research International}, 2016, 2016.

\bibitem{kim2012modeling}
P.S. Kim and P.P. Lee.
\newblock Modeling protective anti-tumor immunity via preventative cancer
  vaccines using a hybrid agent-based and delay differential equation approach.
\newblock {\em PLoS Computational Biology}, 8(10):e1002742, 2012.

\bibitem{shrivastava2018cisplatin}
S.~Shrivastava, U.~Mahantshetty, R.~Engineer, S.~Chopra, R.~Hawaldar, V.~Hande,
  R.A. Kerkar, A.~Maheshwari, T.S. Shylasree, J.~Ghosh, et~al.
\newblock Cisplatin chemoradiotherapy vs radiotherapy in figo stage iiib
  squamous cell carcinoma of the uterine cervix: a randomized clinical trial.
\newblock {\em JAMA Oncology}, 4(4):506--513, 2018.

\bibitem{maletzki2019chemo}
C.~Maletzki, L.~Wiegele, I.~Nassar, J.~Stenzel, and C.~Junghanss.
\newblock Chemo-immunotherapy improves long-term survival in a preclinical
  model of mmr-d-related cancer.
\newblock {\em Journal for Immunotherapy of Cancer}, 7(1):8, 2019.

\bibitem{brown2016chemoimmunotherapy}
J.R. Brown, M.J. Hallek, and J.M. Pagel.
\newblock Chemoimmunotherapy versus targeted treatment in chronic lymphocytic
  leukemia: when, how long, how much, and in which combination?
\newblock {\em American Society of Clinical Oncology Educational Book},
  36:e387--e398, 2016.

\bibitem{hallek2010addition}
M.~Hallek, K.~Fischer, G.~Fingerle-Rowson, A.M. Fink, R.~Busch, J.~Mayer,
  M.~Hensel, G.~Hopfinger, G.~Hess, U.~Von~Gr{\"u}nhagen, et~al.
\newblock Addition of rituximab to fludarabine and cyclophosphamide in patients
  with chronic lymphocytic leukaemia: a randomised, open-label, phase 3 trial.
\newblock {\em The Lancet}, 376(9747):1164--1174, 2010.

\bibitem{galluzzi2014classification}
L~Galluzzi, E~Vacchelli, J~Bravo-San~Pedro, A~Buqu{\'e}, L~Senovilla, E.E.
  Baracco, N.~Bloy, F.~Castoldi, J.~Abastado, P.~Agostinis, et~al.
\newblock Classification of current anticancer immunotherapies.
\newblock {\em Oncotarget}, 5(24):12472, 2014.

\bibitem{sasse2007chemoimmunotherapy}
A.D. Sasse, E.C. Sasse, L.~GO Clark, L.~Ulloa, and O.A.C. Clark.
\newblock Chemoimmunotherapy versus chemotherapy for metastatic malignant
  melanoma.
\newblock {\em Cochrane Database of Systematic Reviews}, (1), 2007.

\bibitem{zhang2014there}
P.~Zhang, M.~Xi, L.~Zhao, J.~Shen, Q.~Li, L.~He, S.~Liu, and M.~Liu.
\newblock Is there a benefit in receiving concurrent chemoradiotherapy for
  elderly patients with inoperable thoracic esophageal squamous cell carcinoma?
\newblock {\em PloS One}, 9(8):e105270, 2014.

\bibitem{binz2015chemovirotherapy}
E.~Binz and U.M. Lauer.
\newblock Chemovirotherapy: Combining chemotherapeutic treatment with oncolytic
  virotherapy.
\newblock {\em Oncolytic Virotherapy}, 4:39, 2015.

\bibitem{advani1998enhancement}
S.J. Advani, G.S. Sibley, P.Y. Song, D.E. Hallahan, Y.~Kataoka, B.~Roizman, and
  R.R Weichselbaum.
\newblock Enhancement of replication of genetically engineered herpes simplex
  viruses by ionizing radiation: a new paradigm for destruction of
  therapeutically intractable tumors.
\newblock {\em Gene Therapy}, 5(2):160, 1998.

\bibitem{chung2002use}
SM~Chung, SJ~Advani, JD~Bradley, Y~Kataoka, K~Vashistha, SY~Yan, J.M. Markert,
  G.Y. Gillespie, R.J. Whitley, B.~Roizman, et~al.
\newblock The use of a genetically engineered herpes simplex virus (r7020) with
  ionizing radiation for experimental hepatoma.
\newblock {\em Gene Therapy}, 9(1):75, 2002.

\bibitem{blank2002replication}
S.V. Blank, S.C. Rubin, G.~Coukos, K.M. Amin, S.M. Albelda, and K.L.
  Molnar-Kimber.
\newblock Replication-selective herpes simplex virus type 1 mutant therapy of
  cervical cancer is enhanced by low-dose radiation.
\newblock {\em Human Gene Therapy}, 13(5):627--639, 2002.

\bibitem{eifel2006chemoradiotherapy}
P.J. Eifel.
\newblock Chemoradiotherapy in the treatment of cervical cancer.
\newblock In {\em Seminars in Radiation Oncology}, volume~16, pages 177--185.
  Elsevier, 2006.

\bibitem{neuner2009chemoradiotherapy}
G.~Neuner, A.~Patel, and M.~Suntharalingam.
\newblock Chemoradiotherapy for esophageal cancer.
\newblock {\em Gastrointestinal Cancer Research: GCR}, 3(2):57, 2009.

\bibitem{shitara2009chemoradiotherapy}
K.~Shitara and K.~Muro.
\newblock Chemoradiotherapy for treatment of esophageal cancer in japan:
  current status and perspectives.
\newblock {\em Gastrointestinal Cancer Research: GCR}, 3(2):66, 2009.

\bibitem{baxi2016trends}
S.S. Baxi, C.~O'Neill, E.J. Sherman, C.L Atoria, N.Y. Lee, D.d~G Pfister, and
  E.~B Elkin.
\newblock Trends in chemoradiation use in elderly patients with head and neck
  cancer: Changing treatment patterns with cetuximab.
\newblock {\em Head \& Neck}, 38(S1):E165--E171, 2016.

\bibitem{sharkey2011cancer}
R.M. Sharkey and D.M. Goldenberg.
\newblock Cancer radioimmunotherapy.
\newblock {\em Immunotherapy}, 3(3):349--370, 2011.

\bibitem{kraeber2016radioimmunotherapy}
F.~Kraeber-Bod{\'e}r{\'e}, J.~Barbet, and J.~Chatal.
\newblock Radioimmunotherapy: from current clinical success to future
  industrial breakthrough?
\newblock {\em Journal of Nuclear Medicine}, 57(3):329--331, 2016.

\bibitem{gopal2003high}
A.K. Gopal, T.A. Gooley, D.G. Maloney, S.H. Petersdorf, J.F. Eary, J.G.
  Rajendran, S.A. Bush, L.D. Durack, J.~Golden, P.J. Martin, et~al.
\newblock High-dose radioimmunotherapy versus conventional high-dose therapy
  and autologous hematopoietic stem cell transplantation for relapsed
  follicular non-hodgkin lymphoma: a multivariable cohort analysis.
\newblock {\em Blood}, 102(7):2351--2357, 2003.

\bibitem{msaouel2009noninvasive}
P~Msaouel, I.~D Iankov, C~Allen, I~Aderca, M.~J Federspiel, D.J. Tindall, J.C.
  Morris, M.~Koutsilieris, S.J. Russell, and E.~Galanis.
\newblock Noninvasive imaging and radiovirotherapy of prostate cancer using an
  oncolytic measles virus expressing the sodium iodide symporter.
\newblock {\em Molecular Therapy}, 17(12):2041--2048, 2009.

\bibitem{li2011oncolytic}
H.~Li, K.~Peng, and S.J. Russell.
\newblock Oncolytic measles virus encoding thyroidal sodium iodide symporter
  for squamous cell cancer of the head and neck radiovirotherapy.
\newblock {\em Human Gene Therapy}, 23(3):295--301, 2011.

\bibitem{dingli2004image}
D.~Dingli, K.~Peng, M.E. Harvey, P.R. Greipp, M.K. O'Connor, R.~Cattaneo, J.C.
  Morris, and S.J. Russell.
\newblock Image-guided radiovirotherapy for multiple myeloma using a
  recombinant measles virus expressing the thyroidal sodium iodide symporter.
\newblock {\em Blood}, 103(5):1641--1646, 2004.

\bibitem{touchefeu2012radiovirotherapy}
Y.~Touchefeu, P.~Franken, and K.~J.~Harrington.
\newblock Radiovirotherapy: principles and prospects in oncology.
\newblock {\em Current Pharmaceutical Design}, 18(22):3313--3320, 2012.

\bibitem{chiocca2002oncolytic}
E.A. Chiocca.
\newblock Oncolytic viruses.
\newblock {\em Nature Reviews Cancer}, 2(12):938, 2002.

\bibitem{spear2000cytotoxicity}
M.A. Spear, F.~Sun, D.J. Eling, E.~Gilpin, T.J. Kipps, E.A. Chiocca, and
  M.~Bouvet.
\newblock Cytotoxicity, apoptosis, and viral replication in tumor cells treated
  with oncolytic ribonucleotide reductase-defective herpes simplex type 1 virus
  (hrr3) combined with ionizing radiation.
\newblock {\em Cancer Gene Therapy}, 7(7):1051, 2000.

\bibitem{jorgensen2001ionizing}
T.J. Jorgensen, S.~Katz, E.K. Wittmack, S.~Varghese, T.~Todo, S.D. Rabkin, and
  R.L. Martuza.
\newblock Ionizing radiation does not alter the antitumor activity of herpes
  simplex virus vector g207 in subcutaneous tumor models of human and murine
  prostate cancer.
\newblock {\em Neoplasia}, 3(5):451--456, 2001.

\bibitem{matlab}
Mathworks.
\newblock \url{http://www.mathworks.com/help/matlab}.
\newblock Accessed: October 2020.

\bibitem{maple}
Differential equations in maple.
\newblock
  \url{https://www.maplesoft.com/applications/view.aspx?sid=103786&view=html}.
\newblock Accessed: October 2020.

\bibitem{Mathematica}
{W}olfram {L}anguage \& {S}ystem {D}ocumentation {C}entre.
\newblock
  \url{https://reference.wolfram.com/language/guide/EquationSolving.html}.
\newblock Accessed: October 2020.

\bibitem{Python}
Python.
\newblock \url{https://www.python.org/}.
\newblock Accessed: October 2020.

\bibitem{Sage}
Sage.
\newblock
  \url{https://doc.sagemath.org/html/en/reference/calculus/sage/calculus/desolvers.html}.
\newblock Accessed: October 2020.

\bibitem{Maxima}
Maxima, a computer algebra system.
\newblock \url{http://maxima.sourceforge.net/}.
\newblock Accessed: October 2020.

\bibitem{Octave}
{GNU} {O}ctave.
\newblock \url{https://www.gnu.org/software/octave/index}.
\newblock Accessed: October 2020.

\bibitem{MathCad}
Mathcad.
\newblock \url{https://www.wikiwand.com/en/Mathcad}.
\newblock Accessed: October 2020.

\bibitem{Julia}
Julia.
\newblock \url{https://julialang.org}.
\newblock Accessed: October 2020.

\bibitem{R}
The {R} project for statistical computing.
\newblock \url{https://www.r-project.org}.
\newblock Accessed: October 2020.

\bibitem{COPASI}
{COPASI}.
\newblock \url{http://copasi.org/}.
\newblock Accessed: November 2020.

\bibitem{CompuSyn}
Compu{S}yn.
\newblock \url{http://www.combosyn.com/}.
\newblock Accessed: November 2020.

\bibitem{CalcuSyn}
Calu{S}yn.
\newblock \url{ http://www.biosoft.com/w/calcusyn.htm}.
\newblock Accessed: November 2020.

\bibitem{he2015timma}
L.~He, K.~Wennerberg, T.~Aittokallio, and J.~Tang.
\newblock {TIMMA-R}: an {R} package for predicting synergistic multi-targeted
  drug combinations in cancer cell lines or patient-derived samples.
\newblock {\em Bioinformatics}, 31(11):1866--1868, 2015.

\bibitem{ianevski2020synergyfinder}
A.~Ianevski, A.K. Giri, and T.~Aittokallio.
\newblock Synergy{F}inder 2.0: visual analytics of multi-drug combination
  synergies.
\newblock {\em Nucleic Acids Research}, 2020.

\bibitem{di2016combenefit}
G.Y. Di~Veroli, D.~Fornari, C.and~Wang, S.~Mollard, J.L. Bramhall, F.M.
  Richards, and D.I. Jodrell.
\newblock Combenefit: an interactive platform for the analysis and
  visualization of drug combinations.
\newblock {\em Bioinformatics}, 32(18):2866--2868, 2016.

\bibitem{URDME}
{URDME}.
\newblock \url{http://urdme.github.io/urdme/}.
\newblock Accessed: January 2021.

\bibitem{COMSOL}
Understand, predict, and optimize physics-based designs and processes with
  {COMSOL} multiphysics.
\newblock \url{https://www.comsol.com/comsol-multiphysics}.
\newblock Accessed: November 2020.

\bibitem{zagidullin2019drugcomb}
B.~Zagidullin, J.~Aldahdooh, S.~Zheng, W.~Wang, Y.~Wang, J.~Saad, A.~Malyutina,
  M.~Jafari, Z.~Tanoli, A.~Pessia, et~al.
\newblock Drugcomb: an integrative cancer drug combination data portal.
\newblock {\em Nucleic Acids Research}, 47(W1):W43--W51, 2019.

\bibitem{combination}
{NCI ALMANAC}: A new tool for research on cancer drug combinations.
\newblock
  \url{https://www.cancer.gov/news-events/cancer-currents-blog/2017/nci-almanac-drug-combinations}.
\newblock Accessed: January 2021.

\bibitem{liu2020drugcombdb}
H.~Liu, W.~Zhang, B.~Zou, J.~Wang, Y.~Deng, and L.~Deng.
\newblock Drugcombdb: a comprehensive database of drug combinations toward the
  discovery of combinatorial therapy.
\newblock {\em Nucleic Acids Research}, 48(D1):D871--D881, 2020.

\bibitem{bai2019cmttdb}
X.~Bai, X.~Yang, L.~Wu, B.~Zuo, J.~Lin, S.~Wang, J.~Bian, X.~Sang, Y.~He,
  Z.~Yang, et~al.
\newblock {CMTT}db: the cancer molecular targeted therapy database.
\newblock {\em Annals of Translational Medicine}, 7(22), 2019.

\bibitem{de2019cancrox}
P.M. de~{\'A}vila, P.C. de~Melo~Bernardo, R.G.T.M. da~Silva, A.~L. Fachin,
  M.~Marins, E.C. Carit{\'a}, et~al.
\newblock {CANCROX}: a cross-species cancer therapy database.
\newblock {\em Database}, 2019, 2019.

\bibitem{yap2010cancer}
K.Y. Yap, A.~Chan, W.K. Chui, and Y.Z. Chen.
\newblock Cancer informatics for the clinician: an interaction database for
  chemotherapy regimens and antiepileptic drugs.
\newblock {\em Seizure}, 19(1):59--67, 2010.

\bibitem{ianevski2020syntoxprofiler}
A.~Ianevski, S.~Timonen, A.~Kononov, T.~Aittokallio, and A.K. Giri.
\newblock Syn{T}ox{P}rofiler: An interactive analysis of drug combination
  synergy, toxicity and efficacy.
\newblock {\em PLoS Computational Biology}, 16(2):e1007604, 2020.

\bibitem{pantziarka2018redo_db}
P.~Pantziarka, C.~Verbaanderd, V.~Sukhatme, I.R. Capistrano, S.~Crispino,
  B.~Gyawali, I.~Rooman, A.M. Van~Nuffel, L.~Meheus, V.P. Sukhatme, et~al.
\newblock Redo\_{DB}: the repurposing drugs in oncology database.
\newblock {\em ecancermedicalscience}, 12, 2018.

\bibitem{kumar2013cancerdr}
R.~Kumar, K.~Chaudhary, S.~Gupta, H.~Singh, S.~Kumar, A.~Gautam, P.~Kapoor, and
  G.P.S. Raghava.
\newblock Cancer{DR}: cancer drug resistance database.
\newblock {\em Scientific Reports}, 3:1445, 2013.

\bibitem{preziosi2003cancer}
L.~Preziosi.
\newblock {\em Cancer modelling and simulation}.
\newblock CRC Press, 2003.

\bibitem{kelley2010theory}
W.G. Kelley and A.C. Peterson.
\newblock {\em The theory of differential equations: classical and
  qualitative}.
\newblock Springer Science \& Business Media, 2010.

\bibitem{brauer2012mathematical}
F.~Brauer and C.~Castillo-Chavez.
\newblock {\em Mathematical models in population biology and epidemiology},
  volume~2.
\newblock Springer, 2012.

\bibitem{Polyanin}
{E}qworld: The world of mathematical equations.
\newblock \url{http://eqworld.ipmnet.ru/en/software.htm}.
\newblock Accessed: October 2020.

\bibitem{Software}
Mathematical software.
\newblock \url{https://www.mat.univie.ac.at/~neum/software.html}.
\newblock Accessed: October 2020.

\bibitem{dimasi2003price}
J.A. DiMasi, R.W. Hansen, and H.G. Grabowski.
\newblock The price of innovation: new estimates of drug development costs.
\newblock {\em Journal of Health Economics}, 22(2):151--185, 2003.

\bibitem{shun2019role}
L.~Shun, X.~Meng, R.~Feng, L.~Yang, L.~Xing, and J.~Yu.
\newblock The role of radiation oncology in immuno-oncology.
\newblock {\em The Oncologist}, 24(Suppl 1):S42--S52, 2019.

\bibitem{TreatmentResistance}
Identifying novel drug combinations to overcome treatment resistance.
\newblock National Cancer Institute,
  \url{https://www.cancer.gov/about-cancer/treatment/research/drug-combo-resistance},
  Accessed May 2019.

\bibitem{tolcher2018improving}
A.W. Tolcher and L.D. Mayer.
\newblock Improving combination cancer therapy: the combiplex{\textregistered}
  development platform.
\newblock {\em Future Oncology}, 14(13):1317--1332, 2018.

\end{thebibliography}
\addcontentsline{toc}{chapter}{Bibliography}


%
%
%
%
%
%
%
%

\medskip
Received xxxx 20xx; revised xxxx 20xx.
\medskip

\end{document}